\documentclass[aps,prb,twocolumn,superscriptaddress,amssymb]{revtex4-2}
\usepackage{amsmath}
\usepackage{amssymb}
\usepackage{amsthm}
\usepackage{amsfonts}
\usepackage{listings}
\usepackage{enumerate}
\usepackage{latexsym}
\usepackage{color}
\usepackage[colorlinks=true,urlcolor=blue,citecolor=blue,linkcolor=blue]{hyperref}
\usepackage{psfrag}
\usepackage{makecell}

\usepackage{bm}
\usepackage{graphicx}
\usepackage{float}
\usepackage{subfigure}

\usepackage{multirow}
\usepackage{tabularx}

\newcommand{\beq}{\begin{equation}}
\newcommand{\eneq}{\end{equation}}

\newcommand{\bal}{\begin{align}}
\newcommand{\eal}{\end{align}}

\input{epsf}

\begin{document}

\tolerance 10000

\newcommand{\vk}{{\bf k}}

\title{Strain Engineering of the Topological Gap and Tunable Chern Numbers in the 2D Kagome Metal-Organic Framework Eu$_2$(C$_6$H$_4$)$_3$}

\author{Jiaxuan Guo}
\email{guojx@stanford.edu}
\affiliation{Department of Applied Physics, Stanford University, Stanford, California 94305, USA}

\author{Simin Nie}
\email{smnie@stanford.edu}
\affiliation{Department of Materials Science and Engineering, Stanford University, Stanford, California 94305, USA}

\author{Fritz B. Prinz}
\email{fprinz@stanford.edu}
\affiliation{Department of Materials Science and Engineering, Stanford University, Stanford, California 94305, USA}
\affiliation{Department of Mechanical Engineering, Stanford University, Stanford, California 94305, USA}

\date{\today}

\begin{abstract}
The quantum anomalous Hall effect carries dissipationless chiral edge currents without an external magnetic field, yet raising its operating temperature and controlling the number of edge channels remain difficult. Using first-principles calculations, we identify the two-dimensional metal-organic kagome ferromagnet Eu$_2$(C$_6$H$_4$)$_3$ as an intrinsic Chern insulator whose topological gap can be enlarged by mechanical strain. The monolayer has a spin-orbit-coupling-induced gap of 72.7 meV that widens to 124.7 meV under $-8\%$ biaxial strain, realizing a quantum anomalous Hall phase with Chern number $\mathcal{C}=-1$ and a single chiral edge state. The local moments come from the half-filled Eu $4f^{7}$ shell, while the gap-opening spin-orbit coupling is carried by Eu $5d$ states hybridized into the carbon kagome bands. The AB-stacked bilayer couples ferromagnetically and accumulates the per-layer Chern numbers, giving $\mathcal{C}=-2$ with two co-propagating chiral channels. An out-of-plane electric field then drives the bilayer through a sequence of topological transitions among $\mathcal{C}=-2$, $-3$, and $-1$, switching the number of edge channels. Together, strain, stacking, and gating give three distinct handles on the gap and the Chern number within a single stoichiometric material.
\end{abstract}

\maketitle

\section{INTRODUCTION}

A two-dimensional insulator that breaks time-reversal symmetry may acquire a nonzero Chern number $\mathcal{C}$, the integral of the Berry curvature of its occupied bands over the Brillouin zone. Bulk-boundary correspondence manifests this integer in directly measurable transport signatures: $|\mathcal{C}|$ chiral edge channels that are immune to backscattering, and a Hall conductance quantized to $\mathcal{C}e^{2}/h$, in principle without an external magnetic field. Haldane established the principle in a honeycomb model whose periodic flux pattern averages to zero \cite{PhysRevLett.61.2015}, thereby decoupling quantized Hall transport from Landau quantization; in the resulting quantum anomalous Hall effect (QAHE), intrinsic magnetization and spin-orbit coupling (SOC) take over the roles played by the external field and the cyclotron gap. Two figures of merit therefore strongly constrain the technological reach of a QAHE material: the bulk topological gap, which helps determine the temperature scale for quantization together with magnetic ordering and disorder, and the Chern number, which fixes the net number of chiral edge channels.

QAHE has been realized in several material platforms, including magnetically doped topological insulators, intrinsic magnetic topological insulators, moir\'{e} Chern systems, and spin-orbit-proximitized graphene-based structures \cite{RevModPhys.95.011002,doi:10.1126/science.1234414,Deng2019QuantumAH,Liu2019RobustAI,Han2024LargeQAHE}. Magnetic doping of (Bi,Sb)$_2$Te$_3$ produced the first zero-field quantized Hall plateaus, but disorder associated with dilute magnetic moments confined full quantization to the millikelvin regime \cite{doi:10.1126/science.1234414,yu2010quantized,chang2013thin,chang2015high,PhysRevLett.115.057206}. Intrinsic magnetic compounds such as MnBi$_2$Te$_4$ subsequently pushed QAH quantization into the kelvin regime \cite{Deng2019QuantumAH,Liu2019RobustAI}, while moir\'{e} heterostructures and spin-orbit-proximitized rhombohedral graphene have provided alternative routes to orbital or interaction-driven Chern insulating states \cite{doi:10.1126/science.aay5533,doi:10.1021/acs.nanolett.1c00696,doi:10.1126/science.aaw3780,li2021quantum,Han2024LargeQAHE}. Despite this progress, quantization survives only at low temperatures, motivating the search for intrinsic two-dimensional Chern insulators with larger topological gaps and robust magnetic order.

A second challenge is controllability. High-Chern-number QAH states are desirable because they provide multiple co-propagating chiral channels and a Hall conductance larger than $e^{2}/h$. Engineered magnetic-topological-insulator multilayers have realized Chern numbers up to $\mathcal{C}=5$ through designed layer sequences \cite{zhao2020tuning}, and thickness engineering in MnBi$_2$Te$_4$ has accessed higher-Chern-number phases in few-layer devices \cite{nwaa089}. Moir\'{e} Chern magnets further offer electrical control of magnetic and topological states \cite{polshyn2020electrical,PhysRevLett.125.227702,chen2021electrically,tschirhart2023intrinsic}. However, these forms of control are often tied to complex heterostructure fabrication, external-field-assisted magnetic alignment, or deep cryogenic operation. It remains desirable to identify a stoichiometric two-dimensional material in which the topological gap itself can be enlarged by a simple external knob, while few-layer structures retain access to multi-channel Chern phases.

Two-dimensional metal-organic frameworks assembled from triphenyl-metal units are a natural place to look. Rare-earth compounds have already been predicted to host the QAHE: monohalides realize intrinsic Chern insulating states, from the high-Curie-temperature LaCl~\cite{Wu2017RareEarthQAH} to ferromagnetic GdCl and GdBr predicted as large-gap two-dimensional quantum anomalous Hall insulators~\cite{PhysRevB.99.035125}, and the experimentally synthesized star-lattice framework Pr$_2$(C$_6$O$_4$Cl$_2$)$_3$ hosts a planar (in-plane-magnetized) quantum anomalous Hall phase with $\mathcal{C}=\pm1$, where noncollinear ligand orbitals open the gap and the sign of $\mathcal{C}$ follows the in-plane magnetization direction~\cite{Pr2QAHE}. In the triphenyl-metal frameworks we study, the organic linkers impose a kagome network on the frontier carbon $p$ orbitals, which supply the Dirac and flat bands that turn into Chern bands once exchange splitting and SOC gap the crossings, while the metal nodes furnish the local moments and spin-orbit-active orbitals that set the gap \cite{PhysRevB.80.113102,PhysRevB.98.205146,PhysRevB.103.014410,Wang2024TopologicalQM,PhysRevLett.110.106804,PhysRevLett.110.196801,PhysRevB.110.235130}. Within this family, the ytterbium framework Yb$_2$(C$_6$H$_4$)$_3$ was recently identified as a large-gap Chern insulator whose Chern number can be scaled by layer stacking and switched by an out-of-plane electric field \cite{YbKagome2026}. We turn to the europium analogue, focusing on the mechanical amplification of its topological gap. As we show below, this enhancement is rooted in the Eu $5d$ states that carry the spin-orbit coupling of the kagome bands.

Using first-principles calculations, we establish Eu$_2$(C$_6$H$_4$)$_3$ as a stable ferromagnetic monolayer with a strain-tunable topological gap. At its relaxed geometry the SOC gap is $72.7$ meV, placing the $\mathcal{C}=-1$ quantum anomalous Hall phase, confirmed by a single chiral edge mode, above liquid-nitrogen temperatures. Under $-8\%$ biaxial compression it widens to $124.7$ meV with the $\mathcal{C}=-1$ topology preserved. Magnetism and spin-orbit coupling have separate microscopic carriers in this material: the half-filled $4f^{7}$ shell provides the large local moments, while the $5d$ states hybridized into the band edges supply the SOC. Strain tunes the topological energy scale by purely mechanical means. Layer stacking and gating provide two additional handles: the AB-stacked bilayer accumulates the per-layer Chern numbers, and an out-of-plane field switches among the resulting high-Chern states. Together these make one material tunable in both its gap and its number of chiral channels.

\section{COMPUTATIONAL METHODS}

We carried out the density functional theory (DFT) calculations with the Vienna \textit{Ab initio} Simulation Package (VASP)~\cite{PhysRevB.47.558,PhysRevB.54.11169}, describing the core-valence interaction by projector augmented-wave (PAW) pseudopotentials~\cite{PhysRevB.50.17953} and exchange and correlation by the Perdew-Burke-Ernzerhof (PBE) generalized-gradient functional~\cite{PhysRevLett.77.3865}. To capture the localized nature of the Eu $4f$ electrons, we applied the rotationally invariant DFT+U approach established by Dudarev~\cite{PhysRevB.57.1505}, adopting an effective Hubbard $U$ value of 8 eV; the topological gap and the band-edge composition reported below are insensitive to this choice over $U_{\mathrm{eff}}=6$--$10$ eV (Sec.~\ref{sec:mono}). For multi-layer configurations, we added the DFT-D3 dispersion correction with Becke-Johnson damping~\cite{Grimme2011EffectOT}. The periodic images along the out-of-plane direction are separated by more than 20~\AA\ of vacuum.

We sampled the Brillouin zone with a $\Gamma$-centered $7\times7\times1$ Monkhorst-Pack grid~\cite{PhysRevB.13.5188} and truncated the plane-wave basis at 500 eV. Both the internal coordinates and the lattice vectors were relaxed until the total energy and the residual forces converged below $10^{-8}$ eV and $10^{-4}$ eV/\AA, respectively.

The strain energy and in-plane stress were obtained from single-point calculations on the fixed-lattice, internally relaxed structures used for the gap calculations. The biaxial two-dimensional modulus was evaluated from a parabolic fit to $E(\varepsilon)$ over $|\varepsilon|\leq4\%$, using $d^{2}E/d\varepsilon^{2}=2A_{0}(C_{11}+C_{12})$, and from a linear fit to the two-dimensional stress $\sigma_{\mathrm{2D}}(\varepsilon)$. The plane-wave convergence of the stress was checked over cutoff energies of 500--800 eV. The magnetocrystalline anisotropy energy under strain was evaluated in self-consistent noncollinear calculations with spin--orbit coupling as $\mathrm{MAE}=E(\mathbf{M}\parallel x)-E(\mathbf{M}\parallel z)$, using $\mathrm{EDIFF}=10^{-6}$ eV. Symmetry was disabled so that the two magnetization directions used the same $k$-point grid, with their integration points matched one by one.

The phonon dispersions were obtained from the finite-displacement method post-processed by the PHONOPY code~\cite{TOGO20151}. \textit{Ab initio} molecular dynamics (AIMD) simulations were carried out on a $2\times2\times1$ supercell within the canonical (NVT) ensemble, spanning 3 ps with a 1-fs time step at 300 K under a Nos\'e-Hoover thermostat~\cite{nose1984unified,hoover1985canonical}.

To estimate the magnetic transition temperature, we computed the magnetic anisotropy energy and the effective Heisenberg exchange couplings from the Liechtenstein-Katsnelson-Antropov-Gubanov (LKAG) formula~\cite{liechtenstein1987local,terasawa2019efficient,PhysRevB.64.174402} via the QuantumATK software~\cite{smidstrup2019quantumatk}, and fed these parameters into atomistic Monte Carlo simulations using the Vampire code~\cite{evans2014atomistic}.

Finally, we characterized the topology and the edge states with a tight-binding model derived from maximally localized Wannier functions (MLWFs), using the WANNIER90 package~\cite{Mostofi2014AnUV}. This topology model spans the C-$p_x$/$p_y$ subspace and is used for the Chern numbers, Wilson loops, and edge spectra. A separate, larger Wannier model was constructed from scalar-relativistic calculations in a C-$p$, Eu-$d$, and Eu-$f$ basis to resolve the origin of the SOC gap. Atomic $\xi\mathbf{L}\cdot\mathbf{S}$ terms were added to this model one orbital manifold at a time, allowing the SOC sectors to be switched independently. For the orbital-resolved SOC test, projection-only Wannier functions preserve the atomic Eu-$d$ labels by avoiding the orbital-sector mixing introduced during maximal localization. The two constructions provide complementary outputs: the MLWF model determines the topology, while the projection-only extended model compares orbital-manifold and strain responses. The semi-infinite chiral edge spectra were then computed via the iterative Green's function algorithm~\cite{Sancho1985HighlyCS} provided by the WANNIERTOOLS suite~\cite{Wu2017WannierToolsAO}. For the field-driven phases, the Chern number was additionally evaluated directly from the Kohn--Sham spinor wavefunctions on uniform $k$-meshes, using the gauge-invariant discretized-Brillouin-zone construction of Fukui, Hatsugai, and Suzuki~\cite{FHS2005}.

\begin{figure}[]
\includegraphics[width=1\linewidth]{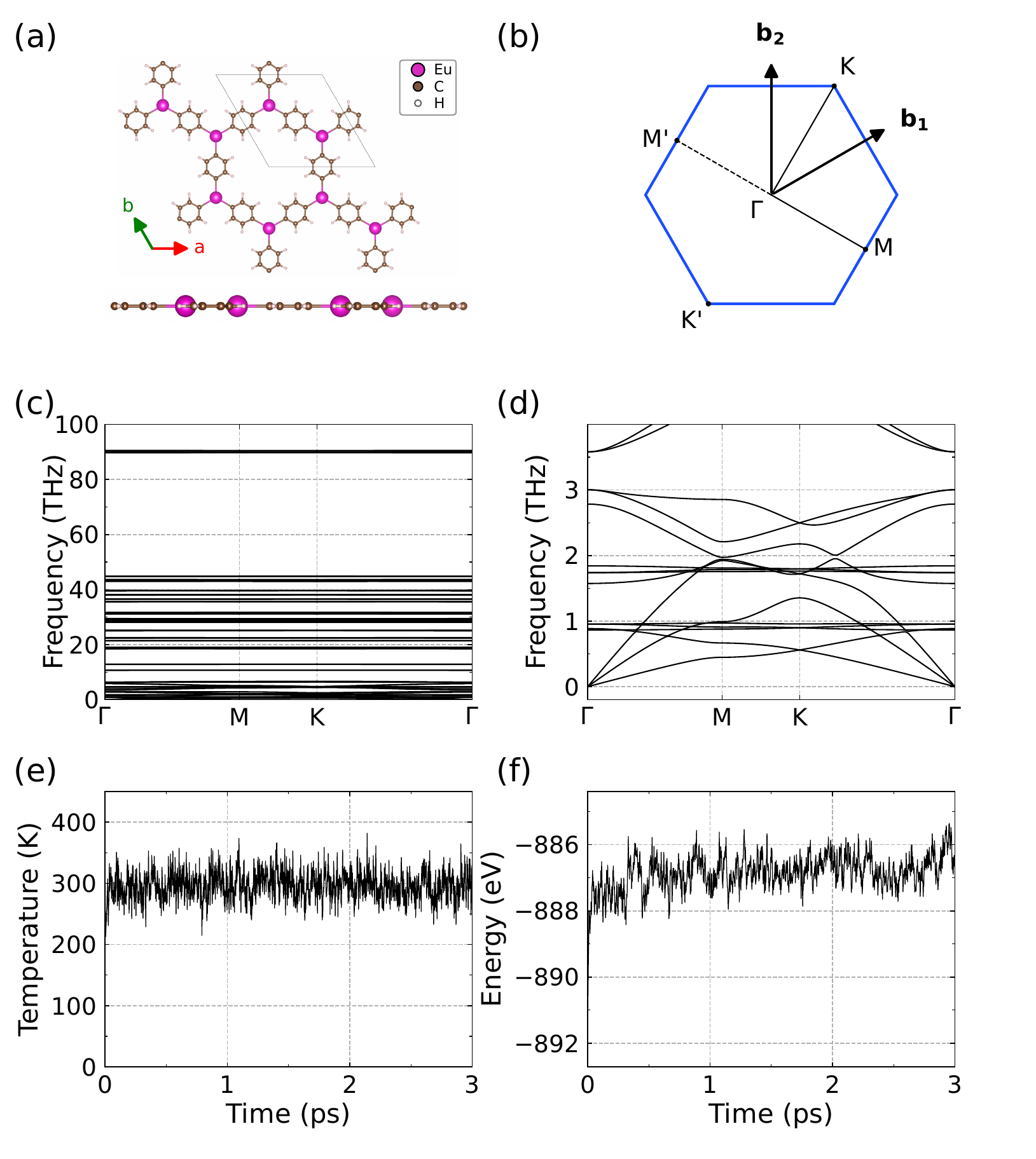}
      \caption{(Color online). (a) Top and side views of Eu$_2$(C$_{6}$H$_{4}$)$_3$. (b) First Brillouin zone and the high-symmetry $k$ points. (c) Phonon dispersion of the monolayer and (d) its low-frequency region. (e),(f) Temperature and total energy during a 3 ps AIMD simulation at 300 K.}
\label{fig1}
\end{figure}

\section{RESULTS}

\subsection{Crystal Structure and Stability}

Figure~\ref{fig1}(a) depicts the atomic configuration of the monolayer Eu$_2$(C$_{6}$H$_{4}$)$_3$. The monolayer crystallizes in the hexagonal space group $P6/m$: each Eu atom sits at the center of a triangle of three benzene rings, and the corner-sharing triangles trace out a 2D metal-organic kagome lattice. After full relaxation the in-plane lattice constant is $a = 13.84$~\AA.

We first evaluated the dynamical stability. The phonon dispersion along the high-symmetry lines, shown in Figs.~\ref{fig1}(c) and \ref{fig1}(d), contains no imaginary branch, so the monolayer is dynamically stable. Thermal stability was then probed by \textit{ab initio} molecular dynamics (AIMD) on a $2\times2\times1$ supercell: over 3 ps at 300 K the framework remains intact with no bond breaking, and the temperature and total energy fluctuate only weakly (Figs.~\ref{fig1}(e) and \ref{fig1}(f)).

In addition to the dynamical and thermal stability evaluations, we assessed the thermodynamic stability of the monolayer with respect to its elemental constituents by computing its formation energy ($E_{\text{form}}$), which is defined according to the following expression:
\begin{eqnarray}
E_{\text{form}} = E[\text{Eu}_{2}(\text{C}_{6}\text{H}_{4})_{3}] - 2\mu[{\text{Eu}}] - 18\mu[{\text{C}}] - 12\mu[{\text{H}}],
\end{eqnarray}
where $E[\text{Eu}_{2}(\text{C}_{6}\text{H}_{4})_{3}]$ denotes the total energy of the primitive unit cell, while $\mu[{\text{Eu}}]$, $\mu[{\text{C}}]$, and $\mu[{\text{H}}]$ represent the chemical potentials of bulk europium, carbon, and hydrogen, respectively. The formation energy is $-9.26$ eV per unit cell ($-0.289$ eV/atom); the monolayer is therefore energetically favored relative to the elemental reference phases.

The layer remains mechanically compliant at the largest compression considered here (Appendix~\ref{app:mechanical_response}). At $\varepsilon=-8\%$, it stores 58.3 meV/atom of strain energy and carries a compressive in-plane two-dimensional stress of only $\approx2.4$ N/m. Independent fits to the energy curvature and stress slope give $C_{11}+C_{12}=19.5$ and 22.4 N/m, respectively, placing $C_{11}+C_{12}$ near 20 N/m, about 20 times smaller than the first-principles value for graphene, $C_{11}+C_{12}=414$--$419$ N/m~\cite{PhysRevB.80.205407,PhysRevB.85.125428}. The bond-resolved response shows that this compliance is accommodated by the soft Eu--C coordination bonds rather than by the rigid phenylene backbone.

\subsection{Intrinsic QAHE and Strain-Engineered Topological Gap}
\label{sec:mono}

Spin-polarized DFT calculations identify a ferromagnetic (FM) ground state for the Eu$_2$(C$_6$H$_4$)$_3$ monolayer, with the easy magnetization axis pointing perpendicular to the lattice plane. The local moments sit predominantly on the Eu sites, accompanied by a sizable induced polarization on the carbon atoms that coordinate directly to europium. To benchmark the magnetic energetics, we evaluated total energies for the FM and antiferromagnetic (AFM) configurations within the same primitive cell. The FM state is energetically preferred by $238.0$ meV over the lowest-energy AFM ordering, indicating strong ferromagnetic exchange. Combining the LKAG-derived Heisenberg couplings with the calculated magnetic anisotropy energy and feeding them into atomistic Monte Carlo simulations yields a Curie temperature of $T_{\text{C}} \approx 116$ K for the monolayer.

\begin{figure}[tp]
\includegraphics[width=1\linewidth]{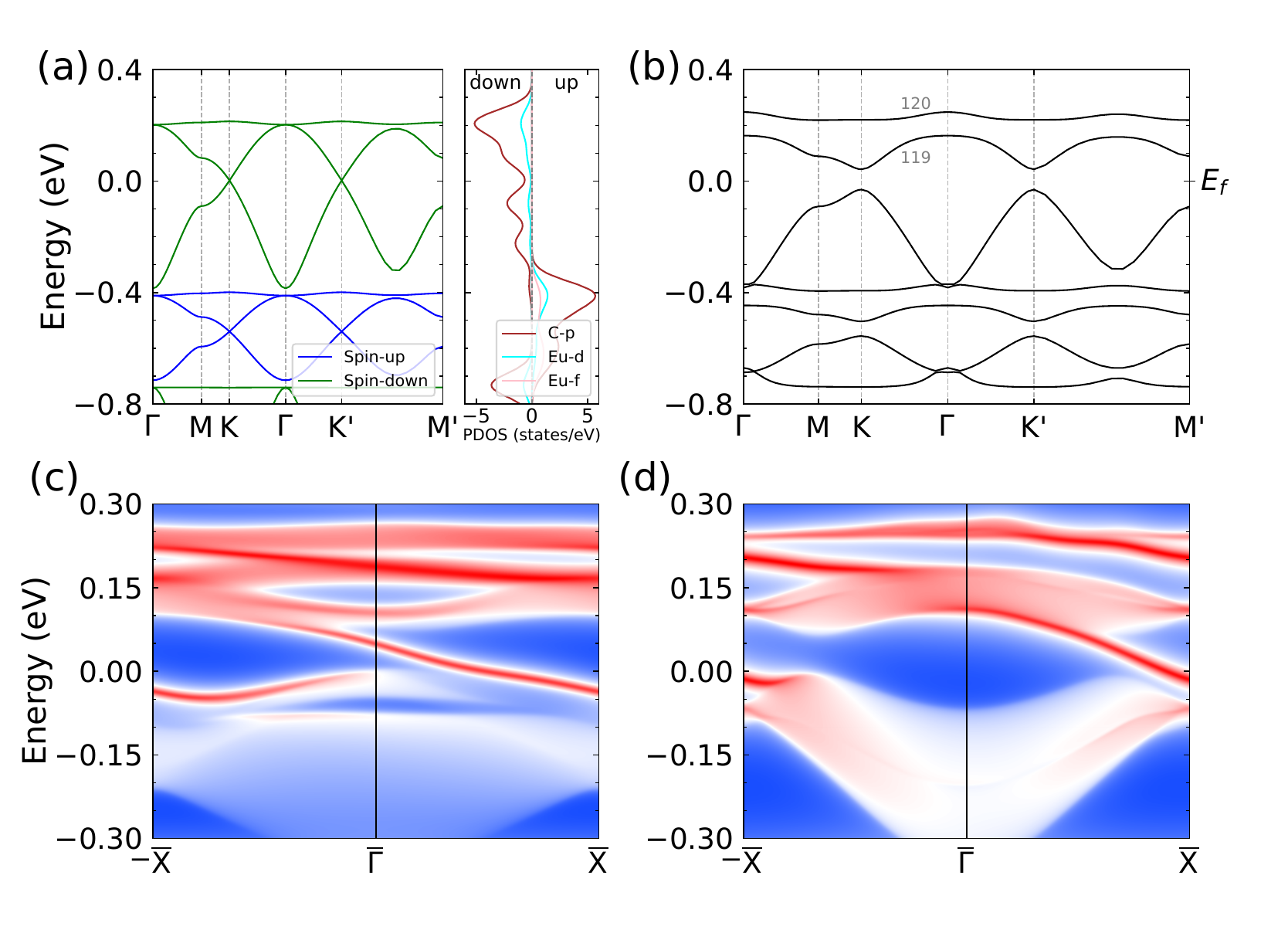}
      \caption{(Color online). The electronic band structures of the monolayer Eu$_2$(C$_{6}$H$_{4}$)$_3$ (a) without SOC, with the projected density of states of the C-$p$, Eu-$d$, and Eu-$f$ orbitals shown alongside, and (b) with SOC, where a 72.7 meV gap opens at K/K$'$. (c)-(d) The armchair and zigzag edge spectra, each showing the single chiral mode of the $\mathcal{C}=-1$ phase.  }
\label{fig2}
\end{figure}

Figure~\ref{fig2}(a) shows the band structure of the FM ground state in the absence of SOC. The bands near the Fermi level belong to a single spin channel and form the characteristic kagome set of two Dirac bands and a flat band~\cite{Kang2019DiracFA,Ye2017MassiveDF,Liu2017GiantAH}; the Dirac crossings sit at the Fermi energy at the K and K$'$ points of the Brillouin zone [Fig.~\ref{fig1}(b)], where they are protected by the $\hat{C}_{3z}$ rotation. The projected density of states shown alongside confirms that this low-energy manifold derives almost entirely from the C-$p$ orbitals, with a minor Eu-$d$ admixture and negligible Eu-$f$ weight. Including SOC lifts the crossings and opens a gap of $72.7$ meV at K/K$'$ [Fig.~\ref{fig2}(b)].

This gap is sizable even though the carbon $p$ orbitals that dominate the low-energy bands carry an intrinsic SOC of only a few meV. Its origin lies in a hybridization channel between the C-$p$ states and the spin-orbit-active orbitals of the metal nodes: even a modest hybridization weight imprints a fraction of the atomic Eu SOC onto the carbon-derived bands, renormalizing the gap to a value far exceeding what the carbon sublattice could produce in isolation. This route to ``borrowed'' SOC underlies the topological gaps of other two-dimensional organometallic frameworks~\cite{PhysRevLett.110.106804,WangOTI2013}. For a rare-earth node the natural candidate donor is the $4f$ shell; the combined projection, symmetry, and manifold-selective tests show, however, that in Eu$_2$(C$_6$H$_4$)$_3$ the borrowed SOC is carried by the Eu $5d$ states.

First, the Eu valence rules out the $4f$ shell at first order. The sphere-projected $4f$ occupation is 6.98 and the local moment is $7.1\,\mu_B$ per Eu (total cell moment $15.9\,\mu_B$ including the polarization induced on the carbon network), identifying Eu$^{2+}$ in the half-filled $4f^{7}$ ($^{8}S_{7/2}$, $L=0$) configuration, whose first-order atomic SOC is quenched; the occupied $4f$ levels lie 0.5--2.5 eV below the Fermi energy. The partially occupied Eu $5d$ shell (occupation 0.43) carries no such constraint. Second, the gap does not track the $4f$ level position. Varying $U_{\mathrm{eff}}$ from 6 to 10 eV shifts the $4f$ manifold by about 2 eV, yet the K/K$'$ gap changes only from 72.5 to 72.7 meV, and the valence, moments, and band-edge composition are unchanged. Third, the orbital-resolved band-edge character at K/K$'$ identifies the Eu weight at the gap edges as $5d$: in the PAW-sphere projections at the valence-band edge (Table~\ref{tab:orbital_character}), the Eu-$5d$ weight is about $11$\% of the total projected weight against about $0.7$\% for Eu-$4f$, a ratio of about $16{:}1$, while the largest share, about $68$\%, resides on the C-$p$ kagome states. Together with the direct tests below, these observations separate the roles of the two Eu shells: the half-filled $4f^{7}$ shell supplies the large local moments, while the $5d$ states hybridized into the band edges supply the SOC that opens the topological gap.

A scalar-relativistic calculation provides the starting point for a direct attribution: without SOC, the two frontier states at K are exactly degenerate, so the entire 72.7 meV gap is opened by SOC. The monolayer has a horizontal mirror symmetry $\sigma_h$, which is preserved by the out-of-plane axial magnetization, and the gap-edge states have even mirror parity. They are formed by C-$(p_x,p_y)$ $\sigma$ states and Eu-$(d_{z^2},d_{x^2-y^2},d_{xy})$ states. The mirror-odd C-$p_z$ and Eu-$(d_{xz},d_{yz})$ weights vanish within numerical precision at both gap edges, whereas an odd-parity control band above the gap has finite weight in both sectors. Within the even-parity Eu-$d$ component, about 90\% of the band-edge weight belongs to the $(d_{x^2-y^2},d_{xy})$ doublet and about 10\% to $d_{z^2}$. The doublet forms the $l_{\mathrm{eff}}=\pm2$ sector, for which $L_zS_z$ is nonzero, whereas $d_{z^2}$ has $L_z=0$.

The extended Wannier Hamiltonian described in the Methods supplies the corresponding causal test (Appendix~\ref{app:soc_origin}). Switching off Eu-$5d$ SOC collapses the gap to zero, whereas switching off Eu-$4f$ or C-$2p$ SOC leaves it unchanged within the reported precision. Retaining SOC only in the Eu-$(d_{x^2-y^2},d_{xy})$ sector recovers 98.8\% of the reference gap; retaining it only in $d_{z^2}$ or $(d_{xz},d_{yz})$ opens no gap. The gap varies linearly with the Eu-$5d$ spin-orbit coupling strength to within a few percent, as a first-order coupling requires (Appendix~\ref{app:soc_origin}). The Eu-$4f$ contribution to the 72.7 meV gap is below 0.05 meV (0.07\%). The empty $4f$ states lie 10--12 eV above $E_F$, and a second-order estimate with this $\approx12$ eV separation between the occupied and empty Eu-$4f$ manifolds in the denominator gives $\approx0.01$ meV. Removing the Eu-$4f$ SOC leaves the gap unchanged, and it remains unchanged when the Eu-$4f$ coupling is increased to five times its atomic value.

To establish the topological character of the gapped phase, we constructed a tight-binding Hamiltonian based on MLWFs spanning the C-$p_x$/$p_y$ subspace. Because the MLWFs are generated from the self-consistent spinor Kohn--Sham states with SOC included, the Eu-$5d$ SOC borrowed by the kagome bands is inherited by the effective C-$p$ hoppings through the downfolding, and the model reproduces the DFT bands in the energy window of interest. Direct integration of the Berry curvature for the occupied manifold over the BZ yields a Chern number $\mathcal{C} = -1$. This assignment is independently confirmed by a Wannier-free diagnostic: evaluating the $\hat{C}_{3z}$ rotational eigenvalues of the occupied bands, read directly from the Kohn--Sham wavefunctions, at $\Gamma$, K and K$'$~\cite{PhysRevB.86.115112,PhysRevLett.108.266802},
\begin{eqnarray}
\exp{(i2\pi \mathcal{C}/3)}=\prod_{n\in occ}(-1)^F\theta_n(\Gamma)\theta_n(K)\theta_n(K'),
\label{eq1}
\end{eqnarray}
where $\theta_n$ are the $\hat{C}_3$ eigenvalues on the $n$-th band and $F=1$ denotes spinful fermions, reproduces $\mathcal{C} \equiv -1\ (\mathrm{mod}\ 3)$ and independently corroborates the Berry-curvature result. The bulk-boundary correspondence is verified in Figs.~\ref{fig2}(c) and \ref{fig2}(d), where iterative Green's function calculations on semi-infinite armchair- and zigzag-terminated ribbons reveal a single chiral mode bridging the bulk valence and conduction continua. The mode disperses with negative group velocity, giving $\mathcal{C} = N_R - N_L = -1$, where $N_R$ ($N_L$) counts the right- (left-)moving modes on a given edge and, throughout this work, $\mathcal{C}$ is defined by integrating the Berry curvature of the occupied bands over the BZ with the surface normal along $+z$, the magnetization direction. These results establish Eu$_2$(C$_6$H$_4$)$_3$ as a monolayer Chern insulator.

\begin{figure}[tp]
\includegraphics[width=1.05\linewidth]{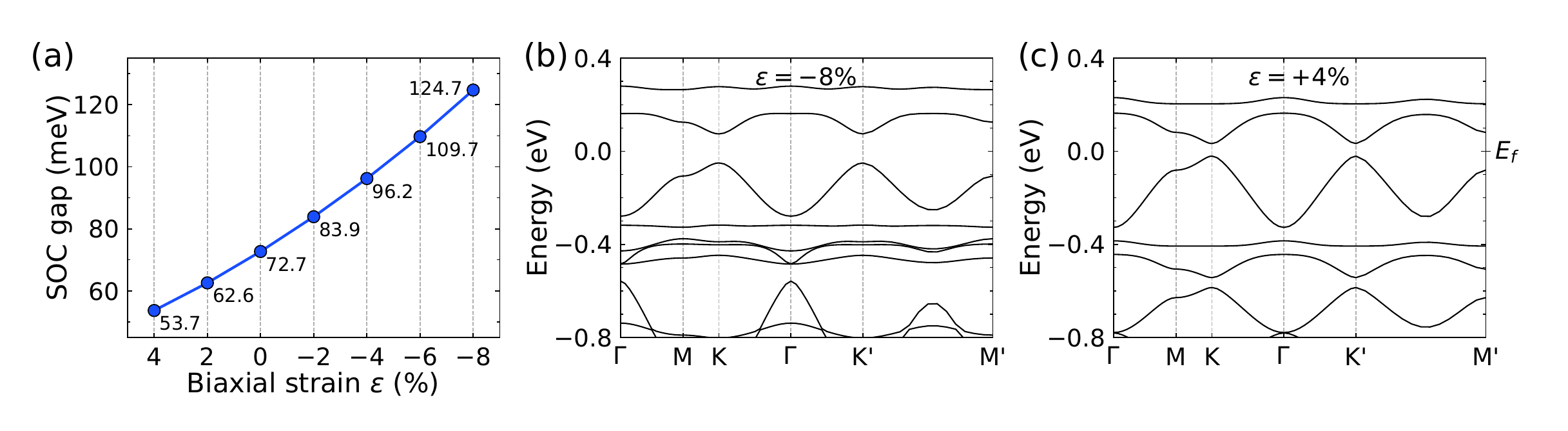}
      \caption{(Color online). (a) Evolution of the SOC-induced topological gap of monolayer Eu$_2$(C$_{6}$H$_{4}$)$_3$ under biaxial strain $\varepsilon$. (b),(c) SOC band structures at $\varepsilon=-8\%$ and $\varepsilon=+4\%$, showing the enlarged and reduced gaps at K/K$'$.  }
\label{fig3}
\end{figure}

Having established the intrinsic QAHE phase, we now turn to the strong response of the topological gap to in-plane lattice deformations. To probe this, we applied biaxial strain $\varepsilon = (a - a_0)/a_0$ in the range $-8\% \leq \varepsilon \leq +4\%$, with $a_0 = 13.84$ \AA\ being the equilibrium lattice constant, and recomputed the SOC-resolved band structure for each strained configuration after a constrained relaxation of the internal coordinates. The resulting gap evolution is summarized in Fig.~\ref{fig3}(a). The gap increases monotonically under compression, rising from $72.7$ meV at $\varepsilon = 0$ to $124.7$ meV at $\varepsilon = -8\%$, an increase of more than $70\%$ relative to the unstrained case. Tensile strain produces the opposite trend, with the gap shrinking but never closing, so that the $\mathcal{C} = -1$ phase is preserved throughout the entire strain window investigated. The magnetism is likewise preserved under this compression: the easy axis remains out of plane, the magnetocrystalline anisotropy energy increases about fourfold (from $\approx2.5$ to $\approx10.3$ meV per cell), and the Curie temperature is $T_{\text{C}}\approx107$ K at $\varepsilon=-8\%$ (compared with $116$ K at $\varepsilon=0$), so the ferromagnetic $\mathcal{C}=-1$ phase stays above liquid-nitrogen temperature across the strain window.

This strain response follows directly from the hybridization-borrowing mechanism identified above. Compressing the lattice contracts the Eu--C coordination distance, increasing the spatial overlap between the carbon $p$ states forming the kagome bands and the Eu $5d$ states that supply the SOC; the spatially extended $5d$ orbitals make this overlap particularly sensitive to the Eu--C distance. The hybridization weight transferred onto the low-energy manifold therefore grows with compression, and so does the effective SOC felt by the band crossings. At $\varepsilon=-8\%$, the Eu--C bonds contract by 9.0\%, more than the 8\% lattice compression, whereas C--C contracts by only 1.5\% and C--H is unchanged. The deformation is therefore absorbed almost entirely by the soft Eu--C coordination bonds, with little change to the phenylene backbone. In the DFT band-edge states, the $(d_{x^2-y^2},d_{xy})$ weight grows by factors of 1.65 at the valence-band edge and 1.68 at the conduction-band edge, while the gap grows by a factor of 1.72, from 72.7 to 124.7 meV. These two independently calculated DFT responses track within about 4\%. With the same $\xi_{5d}$ used for both structures, the extended Wannier model gives a gap ratio of 1.78, compared with the DFT ratio of 1.72, a difference of 3.5\%; the change in hybridization weight therefore accounts for the strain dependence without changing the coupling constant in this construction. As a control, compression raises the band-edge Eu-$4f$ weight from $0.004$ to $0.007$ and strengthens the associated hybridization by about 70\%, yet switching off the Eu-$4f$ SOC still produces no gap change within the reported precision at $-8\%$. Figures~\ref{fig3}(b) and \ref{fig3}(c) show representative band structures at $\varepsilon = -8\%$ and $+4\%$, where the different gap magnitudes at K/K$'$ make the trend apparent.

Moderate compressive strain raises the gap above $120$ meV without leaving the topological phase; this mechanical route requires neither external magnetic fields, nor doping, nor a designed layer sequence. The mechanical cost is correspondingly modest: the layer is about $20\times$ more compliant than graphene, so that $\varepsilon=-8\%$ corresponds to an in-plane stress of only $\approx2.4$ N/m (Appendix~\ref{app:mechanical_response}). We next turn to controlling the Chern number itself through layer stacking and gating.

\subsection{Layer-Dependent Topological Scaling in the Bilayer System}

\begin{figure}[]
\centering
\includegraphics[width=1\linewidth]{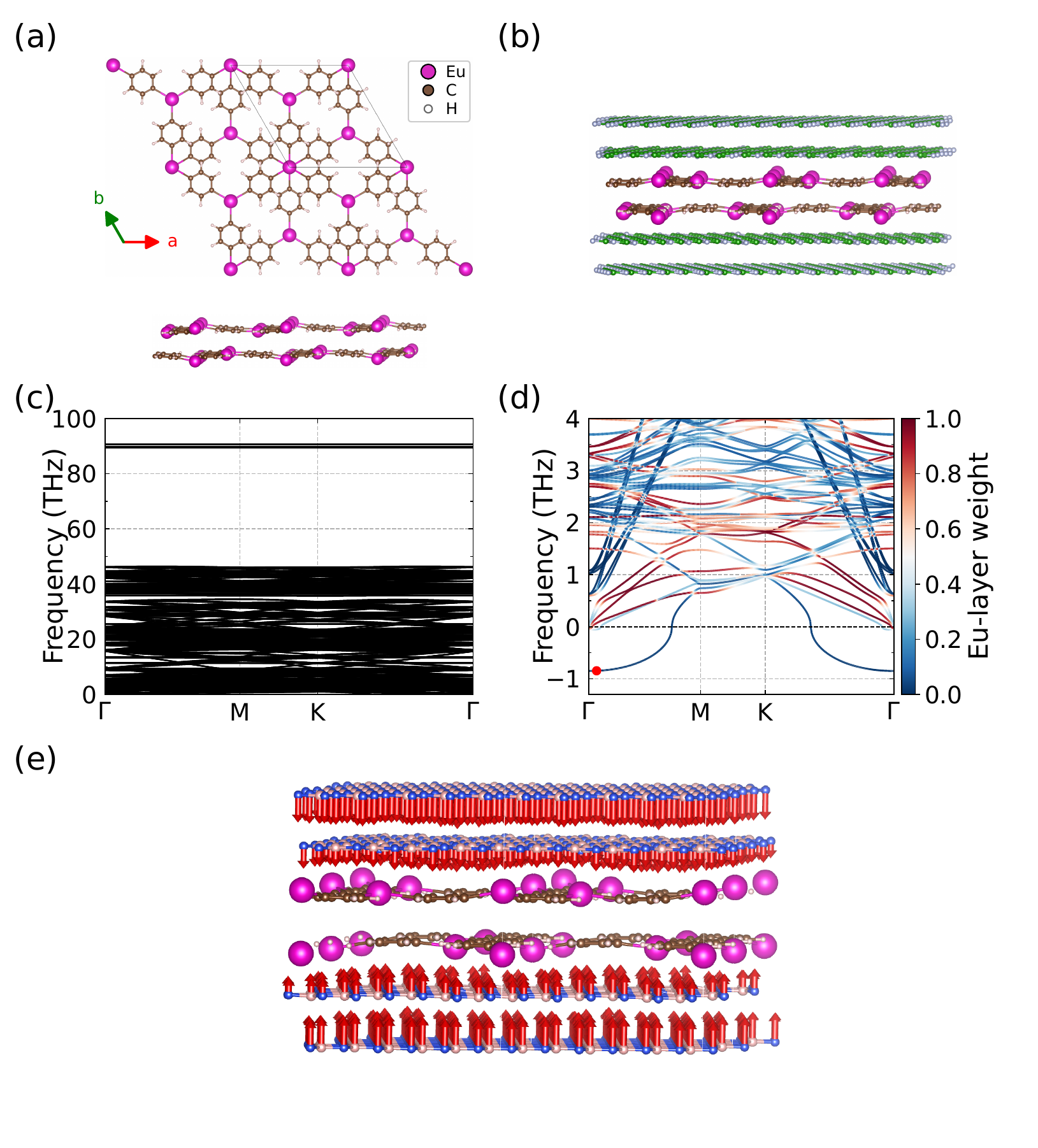}
      \caption{(Color online). (a) Top and side views of the freestanding AB-stacked bilayer Eu$_2$(C$_{6}$H$_{4}$)$_3$. (b) The bilayer encapsulated between hexagonal boron-nitride (h-BN) layers. (c) Phonon spectrum of the h-BN/bilayer/h-BN heterostructure and (d) its low-frequency region color-coded by the phonon weight on the Eu$_2$(C$_{6}$H$_{4}$)$_3$ sublattice (red) versus the h-BN substrate (blue): the single imaginary branch is assigned almost entirely to h-BN displacements, while all bilayer-derived modes are real. (e) Atomic-displacement pattern of this lowest mode at $\Gamma$, with the out-of-plane motion confined to the h-BN layers.}
\label{fig4}
\end{figure}

To scale the Chern number beyond unity we stack two monolayers in an AB arrangement. The stacking lowers the symmetry from the monolayer $P6/m$ to $P\bar{3}$, which retains the $\hat{C}_{3z}$ rotation that the symmetry indicator and the field analysis below rely on, as well as inversion symmetry. After full relaxation the bilayer has an interlayer separation of $3.3$ \r{A} and a small out-of-plane buckling of the two Eu sites within each layer of $0.42$ \r{A}, and the two layers couple ferromagnetically with all moments along the out-of-plane axis. Applying the same LKAG--Monte Carlo protocol as for the monolayer yields a bilayer Curie temperature of $T_{\text{C}} \approx 105$ K, comparable to the monolayer value, so the magnetic order survives stacking. The relaxed AB-stacked geometry is shown in Fig.~\ref{fig4}(a). Encapsulation between hexagonal boron-nitride (h-BN) slabs is a common experimental encapsulation scheme; we adopt this geometry [Fig.~\ref{fig4}(b)], and assess the dynamical stability of the resulting sandwich. The phonon spectrum of the h-BN/bilayer/h-BN heterostructure [Fig.~\ref{fig4}(c)] contains a single imaginary branch, resolved in the low-frequency region [Fig.~\ref{fig4}(d)]; the sublattice-resolved projection there assigns this branch almost entirely to h-BN displacements, while every Eu$_2$(C$_6$H$_4$)$_3$-derived mode remains real, and the displacement pattern of the soft mode [Fig.~\ref{fig4}(e)] confirms that its out-of-plane motion is confined to the h-BN layers. Within the simulated heterostructure, therefore, the single imaginary mode is carried by the h-BN slabs, and no unstable mode dominated by the Eu$_2$(C$_6$H$_4$)$_3$ bilayer is identified.

\begin{figure}[]
\includegraphics[width=1\linewidth]{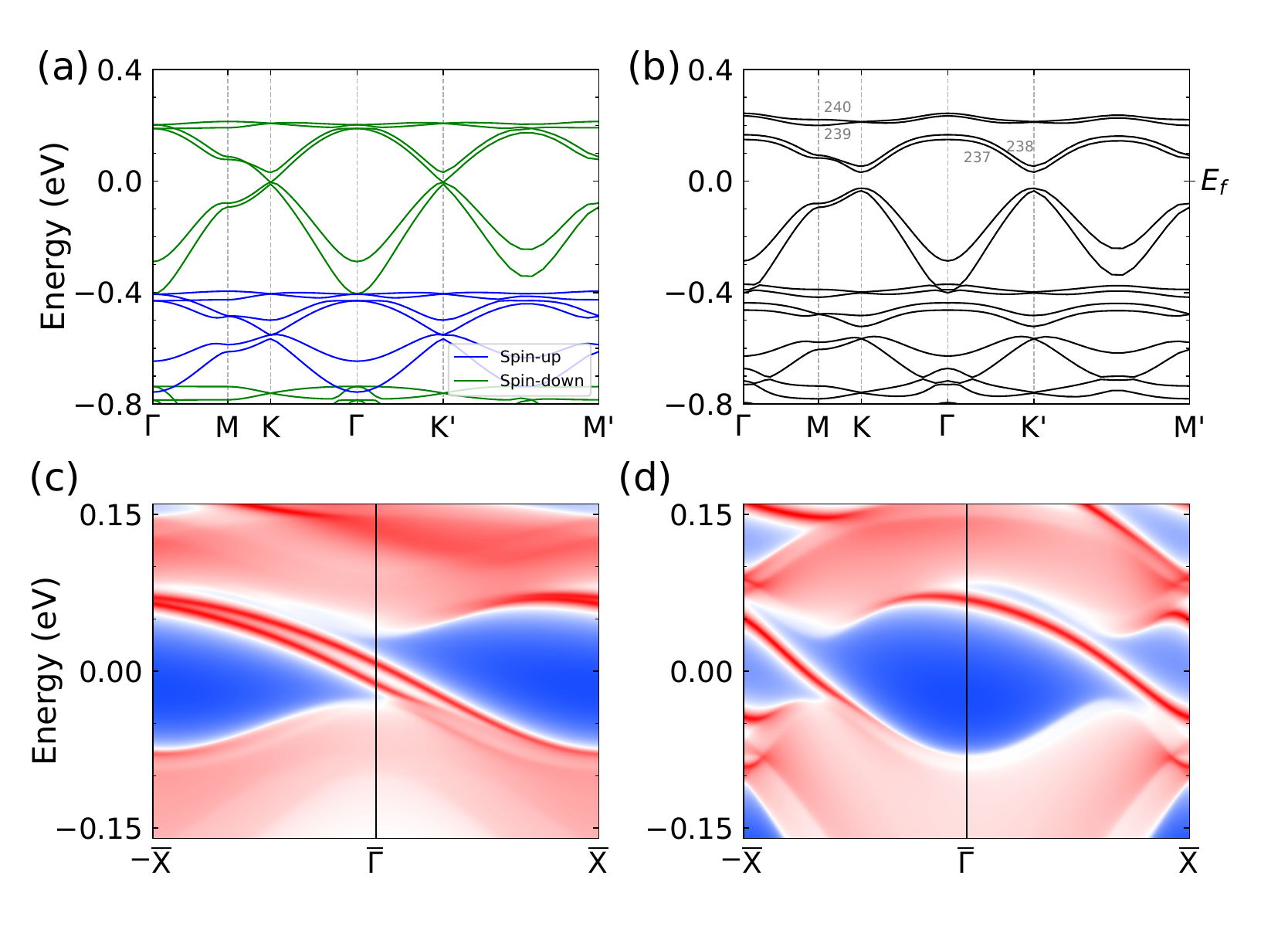}
      \caption{(Color online). Band structures of the bilayer Eu$_2$(C$_{6}$H$_{4}$)$_3$ (a) without and (b) with SOC, where a 57.6 meV gap opens. (c)-(d) The armchair and zigzag edge spectra, each showing the two co-propagating chiral modes of the $\mathcal{C}=-2$ phase.  }
\label{fig5}
\end{figure}

Without SOC, the $\hat{C}_3$-protected band crossings of the two layers sit at the Fermi level [Fig.~\ref{fig5}(a)]; including SOC opens a gap of $57.6$ meV [Fig.~\ref{fig5}(b)]. Applying the $\hat{C}_{3z}$ symmetry indicator of Eq.~(\ref{eq1}) to the four conduction bands just above this gap (Table~\ref{table2}) gives a summed conduction-band Chern number $\equiv2\ (\mathrm{mod}\ 3)$. Because the manifold formed by the occupied bands together with these four conduction bands is separated from the higher bands by a large gap and can be adiabatically connected to a trivial atomic limit, the occupied manifold carries the opposite value, $\mathcal{C}\equiv-2\ (\mathrm{mod}\ 3)$. The Wilson-loop winding of the occupied manifold is $-2$, in agreement with the indicator, which fixes $\mathcal{C}=-2$. The armchair and zigzag edge spectra [Figs.~\ref{fig5}(c) and \ref{fig5}(d)] each show two co-propagating chiral modes dispersing with the same negative group velocity as in the monolayer, $N_R-N_L=-2$, consistent with $\mathcal{C}=-2$ through the bulk-boundary correspondence. The bilayer therefore carries the additive sum of the two monolayer Chern numbers, one chiral channel per layer. As in the monolayer, this gap arises from spin-orbit coupling borrowed onto the carbon kagome bands, so stacking multiplies the chiral channels while each layer retains the same mechanically tunable gap.

\begin{table}[h!]
\centering
\caption{$\hat{C}_{3z}$ eigenvalues at $\Gamma$, K, and K$'$ for the four conduction bands just above the gap of the bilayer Eu$_2$(C$_6$H$_4$)$_3$, together with the Chern number of each band modulo 3 inferred from the product of its rotation eigenvalues via Eq.~(\ref{eq1}).}
\label{table2}
\begin{tabular}{c c c c c}
\hline\hline
Band index & $\Gamma$ & K & K$'$ & $\mathcal{C}_n$ (mod 3) \\ \hline
237 & $e^{-i\pi/3}$ & $e^{i\pi}$    & $e^{i\pi}$    & 1 \\
238 & $e^{-i\pi/3}$ & $e^{i\pi/3}$  & $e^{i\pi/3}$  & 2 \\
239 & $e^{i\pi}$    & $e^{i\pi}$    & $e^{i\pi}$    & 0 \\
240 & $e^{i\pi}$    & $e^{-i\pi/3}$ & $e^{-i\pi/3}$ & 2 \\ \hline\hline
\end{tabular}
\end{table}

\subsection{Electrical Switching of Multi-Channel Edge States}

\begin{figure}[tp]
\includegraphics[width=1\linewidth]{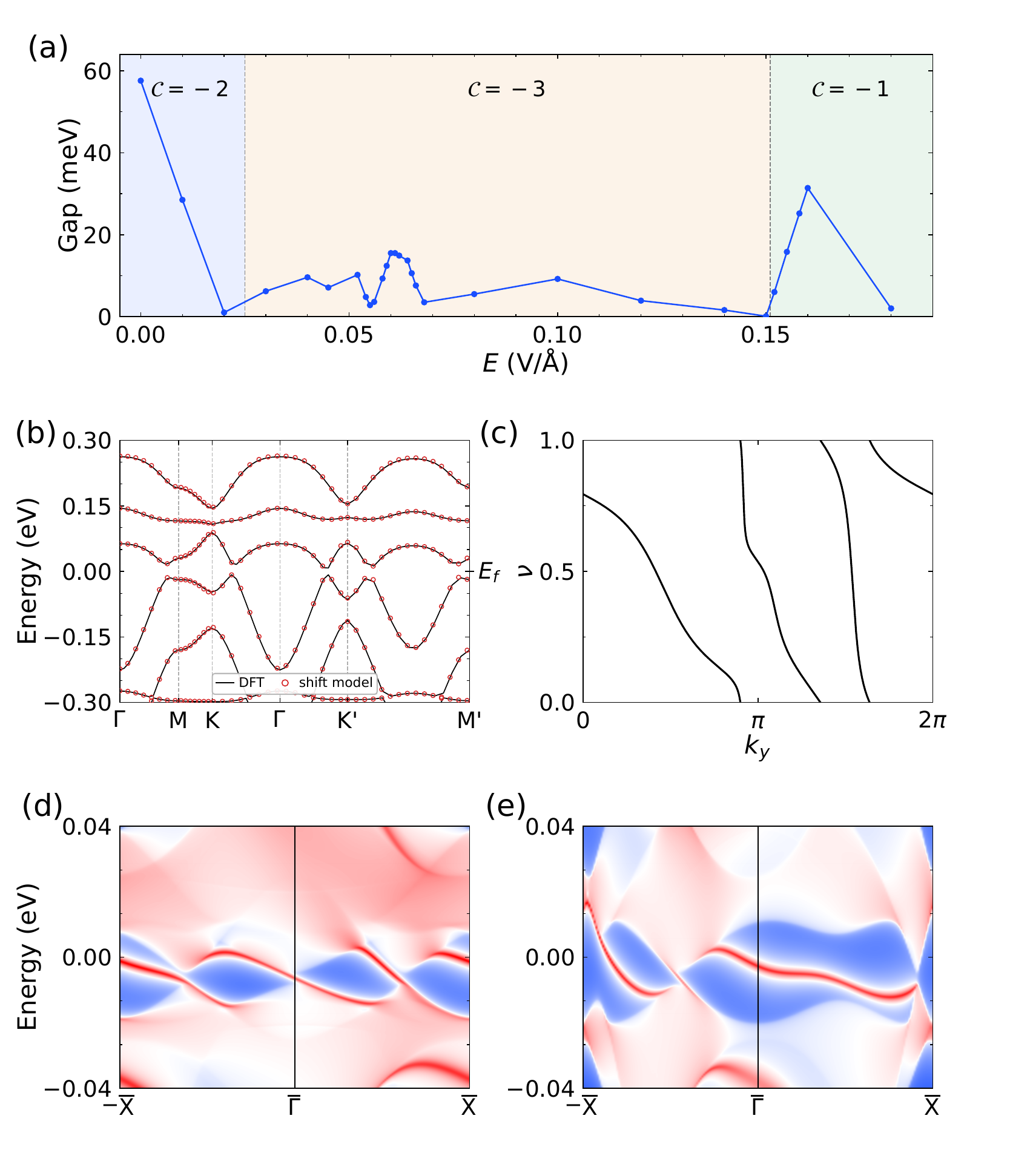}
      \caption{(Color online). (a) Global band gap of the bilayer Eu$_2$(C$_{6}$H$_{4}$)$_3$ as a function of the out-of-plane electric field $E$. Shaded regions mark the three Chern phases $\mathcal{C}=-2$, $-3$, and $-1$ (dashed lines). The first boundary is a gap-closing band inversion at K$'$ and the second one driven by the states at $\Gamma$. (b) Field-on band structure at $E=0.06$~V/\r{A} (black lines), where the $\mathcal{C}=-3$ gap reaches its maximum of $15.5$ meV; open red circles show the field-calibrated Wannier tight-binding bands obtained with a relative interlayer on-site shift $\Delta\varepsilon = 0.20$~eV. (c) Flow of the summed Wannier charge centers $\nu$ of the on-site-shift model in the $\mathcal{C}=-3$ phase; the total winding of $-3$ gives the Chern number. (d),(e) The armchair and zigzag edge spectra of the same model, each showing three co-propagating chiral modes.  }
\label{fig6}
\end{figure}

As in the ytterbium analogue~\cite{YbKagome2026}, an out-of-plane electric field tunes the bilayer Chern number. The field breaks inversion while preserving $\hat{C}_{3z}$, so the symmetry indicator of Eq.~(\ref{eq1}) still applies; but because inversion no longer maps K onto K$'$, the two valleys must be evaluated independently. We therefore read the $\hat{C}_{3z}$ eigenvalues at $\Gamma$, K, and K$'$ separately from the self-consistent field-on wavefunctions, a Wannier-free construction that is insensitive to the field-induced entanglement of the bands. At zero field it returns $\mathcal{C}=-2$, matching the Wilson loop. A weak field closes and reopens the gap at K$'$; beyond $0.03$ V/\r{A} the bilayer carries $\mathcal{C}=-3$, a first transition on a scale comparable to the $0.045$ V/\r{A} of the ytterbium framework.

The $\mathcal{C}=-3$ phase spans a wide window, from $\approx0.03$ to $0.150$ V/\r{A}, and reaches a maximum global gap of $15.5$ meV near $0.06$ V/\r{A}. The symmetry indicator can only fix $\mathcal{C}$ modulo 3, placing this phase in the class of $0$. A direct evaluation of the Chern number from the field-on Kohn--Sham spinors on a discretized Brillouin zone~\cite{FHS2005} gives $\mathcal{C}=-3$ at $0.06$ V/\r{A}; the largest single-plaquette flux is $0.59\pi$, inside the range where the plaquette sum is unambiguous, and an $18\times18$ and a $9\times9$ mesh return the same integer. A second band inversion, driven by the states at $\Gamma$, ends it. The inversion proceeds through a contour rather than through a single point: at $0.150$ V/\r{A} the smallest direct gap on a ring of radius $0.0098$ \r{A}$^{-1}$ about $\Gamma$ is $0.086$ meV, rising to $0.107$ meV where the ring crosses the $\Gamma$--K$'$ line, while the direct gap at $\Gamma$ itself is still $1.18$ meV; the ring contracts with increasing field, from $0.056$ \r{A}$^{-1}$ at $0.140$ V/\r{A}, and closes onto $\Gamma$ at $0.1504$ V/\r{A}. Above that field no such contour survives: radial scans at $0.152$, $0.155$ and $0.158$ V/\r{A} rise monotonically outward from $0.0008$ to $0.15$ \r{A}$^{-1}$ along two inequivalent directions, as a mass term of a single sign requires. The gap then reopens monotonically to a $31$ meV direct gap at $\Gamma$ by $0.16$ V/\r{A}, carrying $\mathcal{C}=-1$. Here the indicator gives $\mathcal{C}\equiv-1\ (\mathrm{mod}\ 3)$, and the same direct evaluation yields $\mathcal{C}=-1$, resolving the mod-3 ambiguity.

The high-field step changes $\mathcal{C}$ by two rather than by one, which a single two-band touching at a $\hat{C}_{3z}$-invariant momentum cannot produce: the two states there differ in rotation eigenvalue by $\Delta l=\pm1$, which admits an interband coupling linear in $k_{\pm}$ and therefore contributes $|\Delta\mathcal{C}|=1$, while $\Delta l=0$ admits a constant coupling and no touching occurs~\cite{PhysRevB.86.115112,PhysRevLett.108.266802}. At $\Gamma$ we read $\Delta l=+1$ directly from the rotation eigenvalues of the two frontier bands. Near such a momentum the two frontier bands are described by $H(\mathbf{k})=d_{z}(\mathbf{k})\sigma_{z}+\mathrm{Re}\,d(\mathbf{k})\,\sigma_{x}+\mathrm{Im}\,d(\mathbf{k})\,\sigma_{y}$, with a gap $2\sqrt{d_{z}^{2}+|d|^{2}}$, so a crossing requires $d_{z}$ and $d$ to vanish at the same point. The selection rule that admits $k_{+}$ also admits $k_{-}^{2}$, so the interband coupling takes the form $d(\mathbf{k})=a\,k_{+}+b\,k_{-}^{2}+\dots$, which vanishes at the invariant momentum and at three further $\hat{C}_{3z}$-related momenta on the shell $|\mathbf{k}|\simeq|a/b|$, while $d_{z}$ is $\hat{C}_{3z}$-invariant and its zeros form a closed contour that contracts as the field grows. At zero field the bilayer is centrosymmetric with $\Gamma$ at the inversion centre, and the two frontier states there carry the same parity, both odd, so $d$ is even in $\mathbf{k}$ and the term linear in $k_{+}$ is forbidden; the field generates it, which is why $|a|$ is small against $|b|$ and the shell sits close to $\Gamma$. For $\Delta l=+1$ the invariant momentum carries winding $+1$ and each satellite $-1$. When $d_{z}$ changes sign at a crossing, the Berry curvature it transfers is its winding times the sense of that change, so the four crossings contribute $-1$ at the centre and $+1$ at each satellite, summing to $\Delta\mathcal{C}=+2$, the value the lattice evaluation gives. The selection rule fixes the windings; the sense of the mass change is what the lattice supplies. The three satellites are $\hat{C}_{3z}$ images of one another and therefore close at a common field; the centre is not related to them by symmetry and need not. Since $a$ is generated by the field, it cannot vanish at a nonzero critical field, so the satellites and the centre close at different fields, and the order is fixed by the contraction reported above: the mass-zero ring reaches $|a/b|$ before it reaches zero, so the three satellites close together first and the invariant momentum last. The shell radius is fixed below at $|a/b|=0.0016$--$0.0033$ \r{A}$^{-1}$, and the same contraction carries the ring from that radius to zero in less than $4\times10^{-5}$ V/\r{A}. The two closures are therefore read as a single step of two on the field grid used here. The coupling is not purely linear in $k_{+}$: such a coupling would leave the gap along the inversion contour flat. On the $0.056$ \r{A}$^{-1}$ contour about $\Gamma$ at $0.140$ V/\r{A} the direct gap runs between $2.235$ and $2.515$ meV, and the two $\hat{C}_{3z}$-equivalent endpoints of that scan differ by $0.006$ meV, forty-six times smaller than that $0.280$ meV range. That modulation fixes the shell radius. Writing the coupling on a contour of radius $r$ as $|d|^{2}=A^{2}+B^{2}+2AB\cos(3\theta+\varphi)$ with $A=ar$ and $B=br^{2}$, the ratio $\rho$ of the smallest to the largest gap on the contour inverts to $|a/b|=r(1-\rho)/(1+\rho)$. The $0.056$ \r{A}$^{-1}$ contour gives $0.0033$ \r{A}$^{-1}$ and the $0.0098$ \r{A}$^{-1}$ contour at $0.150$ V/\r{A} gives $0.0016$ \r{A}$^{-1}$; the factor of two between them is the price of using two terms at radii where the quadratic one already dominates, by a factor $17$ on the outer contour. Either value places the satellite closure within $4\times10^{-5}$ V/\r{A} of the closure at $\Gamma$. The inversion is quadratic in $A/B$ and admits a conjugate root at $r^{2}/|a/b|$, which on the outer contour is $0.95$ \r{A}$^{-1}$, three times $|\Gamma\mathrm{K}'|$ and therefore outside the zone. The counting above needs only that $k_{+}$ and $k_{-}^{2}$ are both admitted, which the rotation eigenvalues fix.

The fields required for both transitions remain within experimentally demonstrated limits. The applied field $E$ in our calculations is the unscreened field of the supercell and therefore corresponds to the displacement field $D/\varepsilon_{0}$ quoted in dual-gate experiments: such devices have opened continuously tunable band gaps of up to $250$ meV in bilayer graphene~\cite{zhang2009direct} and have sustained displacement fields of $2.5$ V/nm ($0.25$ V/\r{A})~\cite{taychatanapat2010electronic}, above the largest field considered here, $0.18$ V/\r{A} ($1.8$ V/nm). The field carried by the gate dielectric itself is smaller by its relative permittivity, $(D/\varepsilon_{0})/\varepsilon_{r}$: for the common gate dielectrics SiO$_2$ and h-BN ($\varepsilon_{r}\approx3.9$ and $\approx3.5$) it reaches at most $\approx0.5$ V/nm across our phase diagram, a fivefold margin below the $2.7$ V/nm irreversible-breakdown strength of ultrathin atomic-layer-deposited SiO$_2$~\cite{usui2013approaching} and a factor of two below the $\approx1.2$ V/nm ($12$ MV/cm) measured for h-BN~\cite{hattori2015layer}, and corresponds to gate voltages of several volts per $10$ nm of dielectric, with the first transition below $0.03$ V/\r{A} requiring only sub-volt biases. The calculated sequence defines the field-accessible topological phase map, while the traversable field range in a specific device is set by its dielectric environment.

The perpendicular field thus drives the bilayer through three distinct Chern phases, $\mathcal{C}=-2\rightarrow-3\rightarrow-1$, each with $|\mathcal{C}|$ chiral edge channels. Figure~\ref{fig6}(a) summarizes this phase diagram together with the gap evolution. To analyze the topology of the field-driven phases, we use a field-calibrated Wannier tight-binding model in which a relative interlayer on-site shift $\Delta\varepsilon$ ($\pm\Delta\varepsilon/2$ per layer) added to the zero-field Wannier Hamiltonian represents the DFT field response; $\Delta\varepsilon$ is fixed by the geometry -- at $E=0.06$~V/\r{A} the potential drops $Ed=0.198$~eV across the $3.3$~\r{A} interlayer separation -- and the model uses $\Delta\varepsilon = 0.20$~eV, which reproduces the $\mathcal{C}=-3$ plateau. Figure~\ref{fig6}(b) validates the on-site-shift model: at $E=0.06$~V/\r{A} the shift-model bands (open circles) are in excellent agreement with the field-on DFT bands (lines) across the Brillouin zone. The validated model then yields the Wilson-loop winding of $-3$ [Fig.~\ref{fig6}(c)] and three co-propagating chiral edge modes with negative group velocity, $N_R-N_L=-3$ [Figs.~\ref{fig6}(d) and \ref{fig6}(e)]. Strain and gating thus act on distinct microscopic levers of the same crystal: strain modulates the Eu--C hybridization that sets the gap scale, while the perpendicular field shifts the interlayer potential balance that selects the Chern number.

\section{CONCLUSION}

Our first-principles calculations establish the two-dimensional metal-organic kagome ferromagnet Eu$_2$(C$_6$H$_4$)$_3$ as an intrinsic Chern insulator with a strain-tunable topological gap. The monolayer is a ferromagnet ($T_{\text{C}}\approx116$ K) hosting a $\mathcal{C}=-1$ quantum anomalous Hall phase with a single chiral edge mode. Its spin-orbit gap of $72.7$ meV widens to $124.7$ meV under $-8\%$ biaxial strain through the borrowing of Eu-$5d$ spin-orbit coupling onto the carbon kagome bands, while the half-filled $4f^{7}$ shell supplies the magnetic order. AB-stacked bilayers couple ferromagnetically and accumulate the layer Chern numbers to $\mathcal{C}=-2$, and an out-of-plane field drives the bilayer through a sequence of topological transitions, $\mathcal{C}=-2\rightarrow-3\rightarrow-1$, that switch the number of chiral channels. Strain thus tunes the size of the topological gap while layer stacking and gating tune the number of chiral channels, making Eu$_2$(C$_6$H$_4$)$_3$ a candidate for mechanically and electrically tunable multi-channel quantum anomalous Hall devices.

\begin{acknowledgments}
We gratefully acknowledge support from the Volkswagen Group of America. F. B. P. received support through a contract with the Volkswagen Group of America. J. G. was supported by the Walecka Fellowship, and S. N. by the Stanford NPL Fund.
\end{acknowledgments}

\section*{Data availability}
The data that support the findings of this article are available from the authors upon
reasonable request.

\appendix

\section{Orbital-resolved origin of the spin-orbit gap}
\label{app:soc_origin}
\setcounter{table}{0}
\renewcommand{\thetable}{A\arabic{table}}
\renewcommand{\theHtable}{A.\arabic{table}}

The scalar-relativistic frontier states are exactly degenerate at K, establishing that the DFT gap is entirely SOC-induced. Table~\ref{tab:orbital_character} resolves the symmetry-allowed hybridization at the two gap edges and compares it with an odd-parity control band.

\begin{table}[H]
\squeezetable
\centering
\setlength{\tabcolsep}{2pt}
\caption{Orbital character at K from the $lm$-resolved PAW projections. The entries are unnormalized PAW-sphere weights. The last row is the total over all ions and channels, which is the denominator for the percentages quoted in the text; the balance to unity resides in interstitial contributions. Channels omitted above (H-$s$, C-$s$, Eu-$s$, Eu-$p$) carry the difference. Band 121 is an odd-parity control band above the gap.}
\label{tab:orbital_character}
\begin{tabular}{lccc}
\hline\hline
Orbital channel & VBM (118) & CBM (119) & Odd control (121) \\ \hline
C-$p_x$ & 0.211 & 0.209 & 0 \\
C-$p_y$ & 0.211 & 0.209 & 0 \\
C-$p_z$ & 0.0000 & 0.0000 & 0.258 \\
Eu-$d_{z^2}$ & 0.007 & 0.006 & 0.000 \\
Eu-$d_{x^2-y^2}$ & 0.029 & 0.033 & 0.000 \\
Eu-$d_{xy}$ & 0.029 & 0.032 & 0.000 \\
Eu-$d_{xz}$ & 0.0000 & 0.0000 & 0.112 \\
Eu-$d_{yz}$ & 0.0000 & 0.0000 & 0.112 \\
Eu-$4f$ & 0.004 & 0.004 & 0.004 \\ \hline
Total & 0.612 & 0.619 & 0.486 \\ \hline\hline
\end{tabular}
\end{table}

At the VBM and CBM, the mirror-odd C-$p_z$ and Eu-$(d_{xz},d_{yz})$ weights vanish within numerical precision, while the control band carries finite weight in both sectors. Normalizing within the even Eu-$d$ component gives about 90\% in the $(d_{x^2-y^2},d_{xy})$ doublet and about 10\% in $d_{z^2}$. The former is the $l_{\mathrm{eff}}=\pm2$ sector with a nonzero $L_zS_z$ matrix element; $d_{z^2}$ has $L_z=0$.

\begin{table}[H]
\centering
\caption{Manifold- and sector-selective SOC test at K for the unstrained structure. Gaps are reported as $\Delta/\Delta_{\mathrm{all}}$, where $\Delta_{\mathrm{all}}$ is the gap with all SOC sectors enabled in the same extended Wannier construction.}
\label{tab:soc_switching}
\begin{tabular}{lc}
\hline\hline
SOC setting & $\Delta/\Delta_{\mathrm{all}}$ \\ \hline
All SOC sectors on & 1.000 \\
All SOC sectors off & 0.000 \\
Eu-$5d$ off & 0.000 \\
Eu-$4f$ off & 1.000 \\
C-$2p$ off & 1.000 \\
Only Eu-$5d$ on & 1.000 \\
Only Eu-$4f$ on & 0.000 \\
Only C-$2p$ on & 0.000 \\
Only Eu-$(d_{x^2-y^2},d_{xy})$ SOC & 0.988 \\
Only Eu-$d_{z^2}$ SOC & 0.000 \\
Only Eu-$(d_{xz},d_{yz})$ SOC & 0.000 \\ \hline\hline
\end{tabular}
\end{table}

With all other inputs fixed, the gap varies linearly with $\xi_{5d}$ to within a few percent, consistent with the first-order relation $\Delta_{\mathrm{SOC}}=2\xi_{5d}w$. Here $\Delta_{\mathrm{SOC}}$ is the SOC-induced gap at K, $\xi_{5d}$ is the atomic Eu-$5d$ spin-orbit coupling constant supplied to the model, and $w$ is the $(d_{x^2-y^2},d_{xy})$ weight of a gap-edge state, normalized over the full Wannier basis so that all orbital weights of that state sum to unity; $w\approx0.09$ for the unstrained structure. This normalization differs from the fractions quoted within the Eu-$d$ component in Sec.~\ref{sec:mono} and from the unnormalized sphere projections of Table~\ref{tab:orbital_character}. The Eu-$4f$ bound, below 0.05 meV or below 0.07\% of the DFT gap, follows from two independent routes: a second-order estimate whose denominator is the $\approx12$ eV separation between the occupied and empty Eu-$4f$ manifolds, which places the empty $4f$ states 10--12 eV above $E_F$; and the zero response of the gap to removal of the Eu-$4f$ SOC, which survives a fivefold increase of $\xi_{4f}$.

This analysis assigns the coupling to Eu-$(d_{x^2-y^2},d_{xy})$ hybridized with C-$(p_x,p_y)$; the assignment does not depend on partitioning SOC into on-site and intersite terms. The atomic-$\xi$ construction provides the relative sector and strain comparisons in Table~\ref{tab:soc_switching}.

\section{Mechanical response under compression}
\label{app:mechanical_response}
\setcounter{table}{0}
\renewcommand{\thetable}{B\arabic{table}}
\renewcommand{\theHtable}{B.\arabic{table}}

Table~\ref{tab:mechanics} collects the strain energy, stress and elastic response at the largest compression considered.

\begin{table}[H]
\squeezetable
\centering
\setlength{\tabcolsep}{1pt}
\caption{Mechanical response at $\varepsilon=-8\%$. The stress rows list the VASP in-plane stress output and the corresponding mechanical two-dimensional stress; negative $\sigma_{\mathrm{2D}}$ denotes compression. The graphene benchmark is obtained from the first-principles elastic constants reported in Refs.~\cite{PhysRevB.80.205407,PhysRevB.85.125428}.}
\label{tab:mechanics}
\begin{tabular}{lcc}
\hline\hline
Quantity or setting & Raw $\sigma_{xx}$ (kbar) & Result \\ \hline
Strain energy & -- & 58.3 meV/atom \\
ENCUT = 500 eV & 12.0068 & $-2.408$ N/m \\
ENCUT = 600 eV & 11.5725 & $-2.321$ N/m \\
ENCUT = 700 eV & 11.6361 & $-2.334$ N/m \\
ENCUT = 800 eV & 11.7223 & $-2.351$ N/m \\
Modulus (energy) & -- & 19.5 N/m \\
Modulus (stress slope) & -- & 22.4 N/m \\
Graphene $C_{11}+C_{12}$ & -- & $414$--$419$ N/m \\ \hline\hline
\end{tabular}
\end{table}

Increasing the plane-wave cutoff from 500 to 800 eV changes the in-plane stress by only 0.28 kbar, or 2.4\% of its magnitude. The finite-window parabolic energy fit and the linear stress fit, which weight anharmonicity differently, yield 19.5 and 22.4 N/m, respectively. Both routes place $C_{11}+C_{12}$ near 20 N/m, quantifying the compliance accommodated by the Eu--C coordination network.

\bibliography{ref}

%apsrev4-2.bst 2019-01-14 (MD) hand-edited version of apsrev4-1.bst
%Control: key (0)
%Control: author (8) initials jnrlst
%Control: editor formatted (1) identically to author
%Control: production of article title (0) allowed
%Control: page (0) single
%Control: year (1) truncated
%Control: production of eprint (0) enabled
\begin{thebibliography}{62}%
\makeatletter
\providecommand \@ifxundefined [1]{%
 \@ifx{#1\undefined}
}%
\providecommand \@ifnum [1]{%
 \ifnum #1\expandafter \@firstoftwo
 \else \expandafter \@secondoftwo
 \fi
}%
\providecommand \@ifx [1]{%
 \ifx #1\expandafter \@firstoftwo
 \else \expandafter \@secondoftwo
 \fi
}%
\providecommand \natexlab [1]{#1}%
\providecommand \enquote  [1]{``#1''}%
\providecommand \bibnamefont  [1]{#1}%
\providecommand \bibfnamefont [1]{#1}%
\providecommand \citenamefont [1]{#1}%
\providecommand \href@noop [0]{\@secondoftwo}%
\providecommand \href [0]{\begingroup \@sanitize@url \@href}%
\providecommand \@href[1]{\@@startlink{#1}\@@href}%
\providecommand \@@href[1]{\endgroup#1\@@endlink}%
\providecommand \@sanitize@url [0]{\catcode `\\12\catcode `\$12\catcode
  `\&12\catcode `\#12\catcode `\^12\catcode `\_12\catcode `\%12\relax}%
\providecommand \@@startlink[1]{}%
\providecommand \@@endlink[0]{}%
\providecommand \url  [0]{\begingroup\@sanitize@url \@url }%
\providecommand \@url [1]{\endgroup\@href {#1}{\urlprefix }}%
\providecommand \urlprefix  [0]{URL }%
\providecommand \Eprint [0]{\href }%
\providecommand \doibase [0]{https://doi.org/}%
\providecommand \selectlanguage [0]{\@gobble}%
\providecommand \bibinfo  [0]{\@secondoftwo}%
\providecommand \bibfield  [0]{\@secondoftwo}%
\providecommand \translation [1]{[#1]}%
\providecommand \BibitemOpen [0]{}%
\providecommand \bibitemStop [0]{}%
\providecommand \bibitemNoStop [0]{.\EOS\space}%
\providecommand \EOS [0]{\spacefactor3000\relax}%
\providecommand \BibitemShut  [1]{\csname bibitem#1\endcsname}%
\let\auto@bib@innerbib\@empty
%</preamble>
\bibitem [{\citenamefont {Haldane}(1988)}]{PhysRevLett.61.2015}%
  \BibitemOpen
  \bibfield  {author} {\bibinfo {author} {\bibfnamefont {F.~D.~M.}\
  \bibnamefont {Haldane}},\ }\bibfield  {title} {\bibinfo {title} {Model for a
  quantum {H}all effect without {L}andau levels: Condensed-matter realization
  of the "parity anomaly"},\ }\href
  {https://doi.org/10.1103/PhysRevLett.61.2015} {\bibfield  {journal} {\bibinfo
   {journal} {Phys. Rev. Lett.}\ }\textbf {\bibinfo {volume} {61}},\ \bibinfo
  {pages} {2015} (\bibinfo {year} {1988})}\BibitemShut {NoStop}%
\bibitem [{\citenamefont {Chang}\ \emph {et~al.}(2023)\citenamefont {Chang},
  \citenamefont {Liu},\ and\ \citenamefont {MacDonald}}]{RevModPhys.95.011002}%
  \BibitemOpen
  \bibfield  {author} {\bibinfo {author} {\bibfnamefont {C.-Z.}\ \bibnamefont
  {Chang}}, \bibinfo {author} {\bibfnamefont {C.-X.}\ \bibnamefont {Liu}},\
  and\ \bibinfo {author} {\bibfnamefont {A.~H.}\ \bibnamefont {MacDonald}},\
  }\bibfield  {title} {\bibinfo {title} {Colloquium: Quantum anomalous {H}all
  effect},\ }\href {https://doi.org/10.1103/RevModPhys.95.011002} {\bibfield
  {journal} {\bibinfo  {journal} {Rev. Mod. Phys.}\ }\textbf {\bibinfo {volume}
  {95}},\ \bibinfo {pages} {011002} (\bibinfo {year} {2023})}\BibitemShut
  {NoStop}%
\bibitem [{\citenamefont {Chang}\ \emph
  {et~al.}(2013{\natexlab{a}})\citenamefont {Chang}, \citenamefont {Zhang},
  \citenamefont {Feng}, \citenamefont {Shen}, \citenamefont {Zhang},
  \citenamefont {Guo}, \citenamefont {Li}, \citenamefont {Ou}, \citenamefont
  {Wei}, \citenamefont {Wang}, \citenamefont {Ji}, \citenamefont {Feng},
  \citenamefont {Ji}, \citenamefont {Chen}, \citenamefont {Jia}, \citenamefont
  {Dai}, \citenamefont {Fang}, \citenamefont {Zhang}, \citenamefont {He},
  \citenamefont {Wang}, \citenamefont {Lu}, \citenamefont {Ma},\ and\
  \citenamefont {Xue}}]{doi:10.1126/science.1234414}%
  \BibitemOpen
  \bibfield  {author} {\bibinfo {author} {\bibfnamefont {C.-Z.}\ \bibnamefont
  {Chang}}, \bibinfo {author} {\bibfnamefont {J.}~\bibnamefont {Zhang}},
  \bibinfo {author} {\bibfnamefont {X.}~\bibnamefont {Feng}}, \bibinfo {author}
  {\bibfnamefont {J.}~\bibnamefont {Shen}}, \bibinfo {author} {\bibfnamefont
  {Z.}~\bibnamefont {Zhang}}, \bibinfo {author} {\bibfnamefont
  {M.}~\bibnamefont {Guo}}, \bibinfo {author} {\bibfnamefont {K.}~\bibnamefont
  {Li}}, \bibinfo {author} {\bibfnamefont {Y.}~\bibnamefont {Ou}}, \bibinfo
  {author} {\bibfnamefont {P.}~\bibnamefont {Wei}}, \bibinfo {author}
  {\bibfnamefont {L.-L.}\ \bibnamefont {Wang}}, \bibinfo {author}
  {\bibfnamefont {Z.-Q.}\ \bibnamefont {Ji}}, \bibinfo {author} {\bibfnamefont
  {Y.}~\bibnamefont {Feng}}, \bibinfo {author} {\bibfnamefont {S.}~\bibnamefont
  {Ji}}, \bibinfo {author} {\bibfnamefont {X.}~\bibnamefont {Chen}}, \bibinfo
  {author} {\bibfnamefont {J.}~\bibnamefont {Jia}}, \bibinfo {author}
  {\bibfnamefont {X.}~\bibnamefont {Dai}}, \bibinfo {author} {\bibfnamefont
  {Z.}~\bibnamefont {Fang}}, \bibinfo {author} {\bibfnamefont {S.-C.}\
  \bibnamefont {Zhang}}, \bibinfo {author} {\bibfnamefont {K.}~\bibnamefont
  {He}}, \bibinfo {author} {\bibfnamefont {Y.}~\bibnamefont {Wang}}, \bibinfo
  {author} {\bibfnamefont {L.}~\bibnamefont {Lu}}, \bibinfo {author}
  {\bibfnamefont {X.-C.}\ \bibnamefont {Ma}},\ and\ \bibinfo {author}
  {\bibfnamefont {Q.-K.}\ \bibnamefont {Xue}},\ }\bibfield  {title} {\bibinfo
  {title} {Experimental observation of the quantum anomalous {H}all effect in a
  magnetic topological insulator},\ }\href
  {https://doi.org/10.1126/science.1234414} {\bibfield  {journal} {\bibinfo
  {journal} {Science}\ }\textbf {\bibinfo {volume} {340}},\ \bibinfo {pages}
  {167} (\bibinfo {year} {2013}{\natexlab{a}})}\BibitemShut {NoStop}%
\bibitem [{\citenamefont {Deng}\ \emph {et~al.}(2020)\citenamefont {Deng},
  \citenamefont {Yu}, \citenamefont {Shi}, \citenamefont {Guo}, \citenamefont
  {Xu}, \citenamefont {Wang}, \citenamefont {Chen},\ and\ \citenamefont
  {Zhang}}]{Deng2019QuantumAH}%
  \BibitemOpen
  \bibfield  {author} {\bibinfo {author} {\bibfnamefont {Y.}~\bibnamefont
  {Deng}}, \bibinfo {author} {\bibfnamefont {Y.}~\bibnamefont {Yu}}, \bibinfo
  {author} {\bibfnamefont {M.~Z.}\ \bibnamefont {Shi}}, \bibinfo {author}
  {\bibfnamefont {Z.}~\bibnamefont {Guo}}, \bibinfo {author} {\bibfnamefont
  {Z.}~\bibnamefont {Xu}}, \bibinfo {author} {\bibfnamefont {J.}~\bibnamefont
  {Wang}}, \bibinfo {author} {\bibfnamefont {X.~H.}\ \bibnamefont {Chen}},\
  and\ \bibinfo {author} {\bibfnamefont {Y.}~\bibnamefont {Zhang}},\ }\bibfield
   {title} {\bibinfo {title} {Quantum anomalous {H}all effect in intrinsic
  magnetic topological insulator {MnBi2Te4}},\ }\href
  {https://doi.org/10.1126/science.aax8156} {\bibfield  {journal} {\bibinfo
  {journal} {Science}\ }\textbf {\bibinfo {volume} {367}},\ \bibinfo {pages}
  {895} (\bibinfo {year} {2020})}\BibitemShut {NoStop}%
\bibitem [{\citenamefont {Liu}\ \emph {et~al.}(2020)\citenamefont {Liu},
  \citenamefont {Wang}, \citenamefont {Li}, \citenamefont {Wu}, \citenamefont
  {Li}, \citenamefont {Li}, \citenamefont {He}, \citenamefont {Xu},
  \citenamefont {Zhang},\ and\ \citenamefont {Wang}}]{Liu2019RobustAI}%
  \BibitemOpen
  \bibfield  {author} {\bibinfo {author} {\bibfnamefont {C.}~\bibnamefont
  {Liu}}, \bibinfo {author} {\bibfnamefont {Y.}~\bibnamefont {Wang}}, \bibinfo
  {author} {\bibfnamefont {H.}~\bibnamefont {Li}}, \bibinfo {author}
  {\bibfnamefont {Y.}~\bibnamefont {Wu}}, \bibinfo {author} {\bibfnamefont
  {Y.}~\bibnamefont {Li}}, \bibinfo {author} {\bibfnamefont {J.}~\bibnamefont
  {Li}}, \bibinfo {author} {\bibfnamefont {K.}~\bibnamefont {He}}, \bibinfo
  {author} {\bibfnamefont {Y.}~\bibnamefont {Xu}}, \bibinfo {author}
  {\bibfnamefont {J.}~\bibnamefont {Zhang}},\ and\ \bibinfo {author}
  {\bibfnamefont {Y.}~\bibnamefont {Wang}},\ }\bibfield  {title} {\bibinfo
  {title} {Robust axion insulator and {C}hern insulator phases in a
  two-dimensional antiferromagnetic topological insulator},\ }\href
  {https://doi.org/10.1038/s41563-019-0573-3} {\bibfield  {journal} {\bibinfo
  {journal} {Nat. Mater.}\ }\textbf {\bibinfo {volume} {19}},\ \bibinfo {pages}
  {522} (\bibinfo {year} {2020})}\BibitemShut {NoStop}%
\bibitem [{\citenamefont {Han}\ \emph {et~al.}(2024)\citenamefont {Han},
  \citenamefont {Lu}, \citenamefont {Yao}, \citenamefont {Yang}, \citenamefont
  {Seo}, \citenamefont {Yoon}, \citenamefont {Watanabe}, \citenamefont
  {Taniguchi}, \citenamefont {Fu}, \citenamefont {Zhang},\ and\ \citenamefont
  {Ju}}]{Han2024LargeQAHE}%
  \BibitemOpen
  \bibfield  {author} {\bibinfo {author} {\bibfnamefont {T.}~\bibnamefont
  {Han}}, \bibinfo {author} {\bibfnamefont {Z.}~\bibnamefont {Lu}}, \bibinfo
  {author} {\bibfnamefont {Y.}~\bibnamefont {Yao}}, \bibinfo {author}
  {\bibfnamefont {J.}~\bibnamefont {Yang}}, \bibinfo {author} {\bibfnamefont
  {J.}~\bibnamefont {Seo}}, \bibinfo {author} {\bibfnamefont {C.}~\bibnamefont
  {Yoon}}, \bibinfo {author} {\bibfnamefont {K.}~\bibnamefont {Watanabe}},
  \bibinfo {author} {\bibfnamefont {T.}~\bibnamefont {Taniguchi}}, \bibinfo
  {author} {\bibfnamefont {L.}~\bibnamefont {Fu}}, \bibinfo {author}
  {\bibfnamefont {F.}~\bibnamefont {Zhang}},\ and\ \bibinfo {author}
  {\bibfnamefont {L.}~\bibnamefont {Ju}},\ }\bibfield  {title} {\bibinfo
  {title} {Large quantum anomalous {H}all effect in spin-orbit proximitized
  rhombohedral graphene},\ }\href {https://doi.org/10.1126/science.adk9749}
  {\bibfield  {journal} {\bibinfo  {journal} {Science}\ }\textbf {\bibinfo
  {volume} {384}},\ \bibinfo {pages} {647} (\bibinfo {year}
  {2024})}\BibitemShut {NoStop}%
\bibitem [{\citenamefont {Yu}\ \emph {et~al.}(2010)\citenamefont {Yu},
  \citenamefont {Zhang}, \citenamefont {Zhang}, \citenamefont {Zhang},
  \citenamefont {Dai},\ and\ \citenamefont {Fang}}]{yu2010quantized}%
  \BibitemOpen
  \bibfield  {author} {\bibinfo {author} {\bibfnamefont {R.}~\bibnamefont
  {Yu}}, \bibinfo {author} {\bibfnamefont {W.}~\bibnamefont {Zhang}}, \bibinfo
  {author} {\bibfnamefont {H.-J.}\ \bibnamefont {Zhang}}, \bibinfo {author}
  {\bibfnamefont {S.-C.}\ \bibnamefont {Zhang}}, \bibinfo {author}
  {\bibfnamefont {X.}~\bibnamefont {Dai}},\ and\ \bibinfo {author}
  {\bibfnamefont {Z.}~\bibnamefont {Fang}},\ }\bibfield  {title} {\bibinfo
  {title} {Quantized anomalous {H}all effect in magnetic topological
  insulators},\ }\href {https://doi.org/10.1126/science.1187485} {\bibfield
  {journal} {\bibinfo  {journal} {Science}\ }\textbf {\bibinfo {volume}
  {329}},\ \bibinfo {pages} {61} (\bibinfo {year} {2010})}\BibitemShut
  {NoStop}%
\bibitem [{\citenamefont {Chang}\ \emph
  {et~al.}(2013{\natexlab{b}})\citenamefont {Chang}, \citenamefont {Zhang},
  \citenamefont {Liu}, \citenamefont {Zhang}, \citenamefont {Feng},
  \citenamefont {Li}, \citenamefont {Wang}, \citenamefont {Chen}, \citenamefont
  {Dai}, \citenamefont {Fang}, \citenamefont {Qi}, \citenamefont {Zhang},
  \citenamefont {Wang}, \citenamefont {He}, \citenamefont {Ma},\ and\
  \citenamefont {Xue}}]{chang2013thin}%
  \BibitemOpen
  \bibfield  {author} {\bibinfo {author} {\bibfnamefont {C.-Z.}\ \bibnamefont
  {Chang}}, \bibinfo {author} {\bibfnamefont {J.}~\bibnamefont {Zhang}},
  \bibinfo {author} {\bibfnamefont {M.}~\bibnamefont {Liu}}, \bibinfo {author}
  {\bibfnamefont {Z.}~\bibnamefont {Zhang}}, \bibinfo {author} {\bibfnamefont
  {X.}~\bibnamefont {Feng}}, \bibinfo {author} {\bibfnamefont {K.}~\bibnamefont
  {Li}}, \bibinfo {author} {\bibfnamefont {L.-L.}\ \bibnamefont {Wang}},
  \bibinfo {author} {\bibfnamefont {X.}~\bibnamefont {Chen}}, \bibinfo {author}
  {\bibfnamefont {X.}~\bibnamefont {Dai}}, \bibinfo {author} {\bibfnamefont
  {Z.}~\bibnamefont {Fang}}, \bibinfo {author} {\bibfnamefont {X.-L.}\
  \bibnamefont {Qi}}, \bibinfo {author} {\bibfnamefont {S.-C.}\ \bibnamefont
  {Zhang}}, \bibinfo {author} {\bibfnamefont {Y.}~\bibnamefont {Wang}},
  \bibinfo {author} {\bibfnamefont {K.}~\bibnamefont {He}}, \bibinfo {author}
  {\bibfnamefont {X.-C.}\ \bibnamefont {Ma}},\ and\ \bibinfo {author}
  {\bibfnamefont {Q.-K.}\ \bibnamefont {Xue}},\ }\bibfield  {title} {\bibinfo
  {title} {Thin films of magnetically doped topological insulator with
  carrier-independent long-range ferromagnetic order},\ }\href
  {https://doi.org/10.1002/adma.201203493} {\bibfield  {journal} {\bibinfo
  {journal} {Adv. Mater.}\ }\textbf {\bibinfo {volume} {25}},\ \bibinfo {pages}
  {1065} (\bibinfo {year} {2013}{\natexlab{b}})}\BibitemShut {NoStop}%
\bibitem [{\citenamefont {Chang}\ \emph
  {et~al.}(2015{\natexlab{a}})\citenamefont {Chang}, \citenamefont {Zhao},
  \citenamefont {Kim}, \citenamefont {Zhang}, \citenamefont {Assaf},
  \citenamefont {Heiman}, \citenamefont {Zhang}, \citenamefont {Liu},
  \citenamefont {Chan},\ and\ \citenamefont {Moodera}}]{chang2015high}%
  \BibitemOpen
  \bibfield  {author} {\bibinfo {author} {\bibfnamefont {C.-Z.}\ \bibnamefont
  {Chang}}, \bibinfo {author} {\bibfnamefont {W.}~\bibnamefont {Zhao}},
  \bibinfo {author} {\bibfnamefont {D.~Y.}\ \bibnamefont {Kim}}, \bibinfo
  {author} {\bibfnamefont {H.}~\bibnamefont {Zhang}}, \bibinfo {author}
  {\bibfnamefont {B.~A.}\ \bibnamefont {Assaf}}, \bibinfo {author}
  {\bibfnamefont {D.}~\bibnamefont {Heiman}}, \bibinfo {author} {\bibfnamefont
  {S.-C.}\ \bibnamefont {Zhang}}, \bibinfo {author} {\bibfnamefont
  {C.}~\bibnamefont {Liu}}, \bibinfo {author} {\bibfnamefont {M.~H.}\
  \bibnamefont {Chan}},\ and\ \bibinfo {author} {\bibfnamefont {J.~S.}\
  \bibnamefont {Moodera}},\ }\bibfield  {title} {\bibinfo {title}
  {High-precision realization of robust quantum anomalous {H}all state in a
  hard ferromagnetic topological insulator},\ }\href
  {https://doi.org/10.1038/nmat4204} {\bibfield  {journal} {\bibinfo  {journal}
  {Nat. Mater.}\ }\textbf {\bibinfo {volume} {14}},\ \bibinfo {pages} {473}
  (\bibinfo {year} {2015}{\natexlab{a}})}\BibitemShut {NoStop}%
\bibitem [{\citenamefont {Chang}\ \emph
  {et~al.}(2015{\natexlab{b}})\citenamefont {Chang}, \citenamefont {Zhao},
  \citenamefont {Kim}, \citenamefont {Wei}, \citenamefont {Jain}, \citenamefont
  {Liu}, \citenamefont {Chan},\ and\ \citenamefont
  {Moodera}}]{PhysRevLett.115.057206}%
  \BibitemOpen
  \bibfield  {author} {\bibinfo {author} {\bibfnamefont {C.-Z.}\ \bibnamefont
  {Chang}}, \bibinfo {author} {\bibfnamefont {W.}~\bibnamefont {Zhao}},
  \bibinfo {author} {\bibfnamefont {D.~Y.}\ \bibnamefont {Kim}}, \bibinfo
  {author} {\bibfnamefont {P.}~\bibnamefont {Wei}}, \bibinfo {author}
  {\bibfnamefont {J.~K.}\ \bibnamefont {Jain}}, \bibinfo {author}
  {\bibfnamefont {C.}~\bibnamefont {Liu}}, \bibinfo {author} {\bibfnamefont
  {M.~H.~W.}\ \bibnamefont {Chan}},\ and\ \bibinfo {author} {\bibfnamefont
  {J.~S.}\ \bibnamefont {Moodera}},\ }\bibfield  {title} {\bibinfo {title}
  {Zero-field dissipationless chiral edge transport and the nature of
  dissipation in the quantum anomalous {H}all state},\ }\href
  {https://doi.org/10.1103/PhysRevLett.115.057206} {\bibfield  {journal}
  {\bibinfo  {journal} {Phys. Rev. Lett.}\ }\textbf {\bibinfo {volume} {115}},\
  \bibinfo {pages} {057206} (\bibinfo {year} {2015}{\natexlab{b}})}\BibitemShut
  {NoStop}%
\bibitem [{\citenamefont {Serlin}\ \emph {et~al.}(2020)\citenamefont {Serlin},
  \citenamefont {Tschirhart}, \citenamefont {Polshyn}, \citenamefont {Zhang},
  \citenamefont {Zhu}, \citenamefont {Watanabe}, \citenamefont {Taniguchi},
  \citenamefont {Balents},\ and\ \citenamefont
  {Young}}]{doi:10.1126/science.aay5533}%
  \BibitemOpen
  \bibfield  {author} {\bibinfo {author} {\bibfnamefont {M.}~\bibnamefont
  {Serlin}}, \bibinfo {author} {\bibfnamefont {C.~L.}\ \bibnamefont
  {Tschirhart}}, \bibinfo {author} {\bibfnamefont {H.}~\bibnamefont {Polshyn}},
  \bibinfo {author} {\bibfnamefont {Y.}~\bibnamefont {Zhang}}, \bibinfo
  {author} {\bibfnamefont {J.}~\bibnamefont {Zhu}}, \bibinfo {author}
  {\bibfnamefont {K.}~\bibnamefont {Watanabe}}, \bibinfo {author}
  {\bibfnamefont {T.}~\bibnamefont {Taniguchi}}, \bibinfo {author}
  {\bibfnamefont {L.}~\bibnamefont {Balents}},\ and\ \bibinfo {author}
  {\bibfnamefont {A.~F.}\ \bibnamefont {Young}},\ }\bibfield  {title} {\bibinfo
  {title} {Intrinsic quantized anomalous {H}all effect in a moiré
  heterostructure},\ }\href {https://doi.org/10.1126/science.aay5533}
  {\bibfield  {journal} {\bibinfo  {journal} {Science}\ }\textbf {\bibinfo
  {volume} {367}},\ \bibinfo {pages} {900} (\bibinfo {year}
  {2020})}\BibitemShut {NoStop}%
\bibitem [{\citenamefont {Sharpe}\ \emph {et~al.}(2021)\citenamefont {Sharpe},
  \citenamefont {Fox}, \citenamefont {Barnard}, \citenamefont {Finney},
  \citenamefont {Watanabe}, \citenamefont {Taniguchi}, \citenamefont
  {Kastner},\ and\ \citenamefont
  {Goldhaber-Gordon}}]{doi:10.1021/acs.nanolett.1c00696}%
  \BibitemOpen
  \bibfield  {author} {\bibinfo {author} {\bibfnamefont {A.~L.}\ \bibnamefont
  {Sharpe}}, \bibinfo {author} {\bibfnamefont {E.~J.}\ \bibnamefont {Fox}},
  \bibinfo {author} {\bibfnamefont {A.~W.}\ \bibnamefont {Barnard}}, \bibinfo
  {author} {\bibfnamefont {J.}~\bibnamefont {Finney}}, \bibinfo {author}
  {\bibfnamefont {K.}~\bibnamefont {Watanabe}}, \bibinfo {author}
  {\bibfnamefont {T.}~\bibnamefont {Taniguchi}}, \bibinfo {author}
  {\bibfnamefont {M.~A.}\ \bibnamefont {Kastner}},\ and\ \bibinfo {author}
  {\bibfnamefont {D.}~\bibnamefont {Goldhaber-Gordon}},\ }\bibfield  {title}
  {\bibinfo {title} {Evidence of orbital ferromagnetism in twisted bilayer
  graphene aligned to hexagonal boron nitride},\ }\href
  {https://doi.org/10.1021/acs.nanolett.1c00696} {\bibfield  {journal}
  {\bibinfo  {journal} {Nano Lett.}\ }\textbf {\bibinfo {volume} {21}},\
  \bibinfo {pages} {4299} (\bibinfo {year} {2021})},\ \bibinfo {note} {pMID:
  33970644}\BibitemShut {NoStop}%
\bibitem [{\citenamefont {Sharpe}\ \emph {et~al.}(2019)\citenamefont {Sharpe},
  \citenamefont {Fox}, \citenamefont {Barnard}, \citenamefont {Finney},
  \citenamefont {Watanabe}, \citenamefont {Taniguchi}, \citenamefont
  {Kastner},\ and\ \citenamefont
  {Goldhaber-Gordon}}]{doi:10.1126/science.aaw3780}%
  \BibitemOpen
  \bibfield  {author} {\bibinfo {author} {\bibfnamefont {A.~L.}\ \bibnamefont
  {Sharpe}}, \bibinfo {author} {\bibfnamefont {E.~J.}\ \bibnamefont {Fox}},
  \bibinfo {author} {\bibfnamefont {A.~W.}\ \bibnamefont {Barnard}}, \bibinfo
  {author} {\bibfnamefont {J.}~\bibnamefont {Finney}}, \bibinfo {author}
  {\bibfnamefont {K.}~\bibnamefont {Watanabe}}, \bibinfo {author}
  {\bibfnamefont {T.}~\bibnamefont {Taniguchi}}, \bibinfo {author}
  {\bibfnamefont {M.~A.}\ \bibnamefont {Kastner}},\ and\ \bibinfo {author}
  {\bibfnamefont {D.}~\bibnamefont {Goldhaber-Gordon}},\ }\bibfield  {title}
  {\bibinfo {title} {Emergent ferromagnetism near three-quarters filling in
  twisted bilayer graphene},\ }\href {https://doi.org/10.1126/science.aaw3780}
  {\bibfield  {journal} {\bibinfo  {journal} {Science}\ }\textbf {\bibinfo
  {volume} {365}},\ \bibinfo {pages} {605} (\bibinfo {year}
  {2019})}\BibitemShut {NoStop}%
\bibitem [{\citenamefont {Li}\ \emph {et~al.}(2021)\citenamefont {Li},
  \citenamefont {Jiang}, \citenamefont {Shen}, \citenamefont {Zhang},
  \citenamefont {Li}, \citenamefont {Tao}, \citenamefont {Devakul},
  \citenamefont {Watanabe}, \citenamefont {Taniguchi}, \citenamefont {Fu},
  \citenamefont {Shan},\ and\ \citenamefont {Mak}}]{li2021quantum}%
  \BibitemOpen
  \bibfield  {author} {\bibinfo {author} {\bibfnamefont {T.}~\bibnamefont
  {Li}}, \bibinfo {author} {\bibfnamefont {S.}~\bibnamefont {Jiang}}, \bibinfo
  {author} {\bibfnamefont {B.}~\bibnamefont {Shen}}, \bibinfo {author}
  {\bibfnamefont {Y.}~\bibnamefont {Zhang}}, \bibinfo {author} {\bibfnamefont
  {L.}~\bibnamefont {Li}}, \bibinfo {author} {\bibfnamefont {Z.}~\bibnamefont
  {Tao}}, \bibinfo {author} {\bibfnamefont {T.}~\bibnamefont {Devakul}},
  \bibinfo {author} {\bibfnamefont {K.}~\bibnamefont {Watanabe}}, \bibinfo
  {author} {\bibfnamefont {T.}~\bibnamefont {Taniguchi}}, \bibinfo {author}
  {\bibfnamefont {L.}~\bibnamefont {Fu}}, \bibinfo {author} {\bibfnamefont
  {J.}~\bibnamefont {Shan}},\ and\ \bibinfo {author} {\bibfnamefont {K.~F.}\
  \bibnamefont {Mak}},\ }\bibfield  {title} {\bibinfo {title} {Quantum
  anomalous {H}all effect from intertwined moir{\'e} bands},\ }\href
  {https://doi.org/10.1038/s41586-021-04171-1} {\bibfield  {journal} {\bibinfo
  {journal} {Nature}\ }\textbf {\bibinfo {volume} {600}},\ \bibinfo {pages}
  {641} (\bibinfo {year} {2021})}\BibitemShut {NoStop}%
\bibitem [{\citenamefont {Zhao}\ \emph {et~al.}(2020)\citenamefont {Zhao},
  \citenamefont {Zhang}, \citenamefont {Mei}, \citenamefont {Zhou},
  \citenamefont {Yi}, \citenamefont {Zhang}, \citenamefont {Yu}, \citenamefont
  {Xiao}, \citenamefont {Wang}, \citenamefont {Samarth}, \citenamefont {Chan},
  \citenamefont {Liu},\ and\ \citenamefont {Chang}}]{zhao2020tuning}%
  \BibitemOpen
  \bibfield  {author} {\bibinfo {author} {\bibfnamefont {Y.-F.}\ \bibnamefont
  {Zhao}}, \bibinfo {author} {\bibfnamefont {R.}~\bibnamefont {Zhang}},
  \bibinfo {author} {\bibfnamefont {R.}~\bibnamefont {Mei}}, \bibinfo {author}
  {\bibfnamefont {L.-J.}\ \bibnamefont {Zhou}}, \bibinfo {author}
  {\bibfnamefont {H.}~\bibnamefont {Yi}}, \bibinfo {author} {\bibfnamefont
  {Y.-Q.}\ \bibnamefont {Zhang}}, \bibinfo {author} {\bibfnamefont
  {J.}~\bibnamefont {Yu}}, \bibinfo {author} {\bibfnamefont {R.}~\bibnamefont
  {Xiao}}, \bibinfo {author} {\bibfnamefont {K.}~\bibnamefont {Wang}}, \bibinfo
  {author} {\bibfnamefont {N.}~\bibnamefont {Samarth}}, \bibinfo {author}
  {\bibfnamefont {M.~H.~W.}\ \bibnamefont {Chan}}, \bibinfo {author}
  {\bibfnamefont {C.-X.}\ \bibnamefont {Liu}},\ and\ \bibinfo {author}
  {\bibfnamefont {C.-Z.}\ \bibnamefont {Chang}},\ }\bibfield  {title} {\bibinfo
  {title} {Tuning the {C}hern number in quantum anomalous {H}all insulators},\
  }\href {https://doi.org/10.1038/s41586-020-3020-3} {\bibfield  {journal}
  {\bibinfo  {journal} {Nature}\ }\textbf {\bibinfo {volume} {588}},\ \bibinfo
  {pages} {419} (\bibinfo {year} {2020})}\BibitemShut {NoStop}%
\bibitem [{\citenamefont {Ge}\ \emph {et~al.}(2020)\citenamefont {Ge},
  \citenamefont {Liu}, \citenamefont {Li}, \citenamefont {Li}, \citenamefont
  {Luo}, \citenamefont {Wu}, \citenamefont {Xu},\ and\ \citenamefont
  {Wang}}]{nwaa089}%
  \BibitemOpen
  \bibfield  {author} {\bibinfo {author} {\bibfnamefont {J.}~\bibnamefont
  {Ge}}, \bibinfo {author} {\bibfnamefont {Y.}~\bibnamefont {Liu}}, \bibinfo
  {author} {\bibfnamefont {J.}~\bibnamefont {Li}}, \bibinfo {author}
  {\bibfnamefont {H.}~\bibnamefont {Li}}, \bibinfo {author} {\bibfnamefont
  {T.}~\bibnamefont {Luo}}, \bibinfo {author} {\bibfnamefont {Y.}~\bibnamefont
  {Wu}}, \bibinfo {author} {\bibfnamefont {Y.}~\bibnamefont {Xu}},\ and\
  \bibinfo {author} {\bibfnamefont {J.}~\bibnamefont {Wang}},\ }\bibfield
  {title} {\bibinfo {title} {High-{C}hern-number and high-temperature quantum
  {H}all effect without {L}andau levels},\ }\href
  {https://doi.org/10.1093/nsr/nwaa089} {\bibfield  {journal} {\bibinfo
  {journal} {Natl. Sci. Rev.}\ }\textbf {\bibinfo {volume} {7}},\ \bibinfo
  {pages} {1280} (\bibinfo {year} {2020})}\BibitemShut {NoStop}%
\bibitem [{\citenamefont {Polshyn}\ \emph {et~al.}(2020)\citenamefont
  {Polshyn}, \citenamefont {Zhu}, \citenamefont {Kumar}, \citenamefont {Zhang},
  \citenamefont {Yang}, \citenamefont {Tschirhart}, \citenamefont {Serlin},
  \citenamefont {Watanabe}, \citenamefont {Taniguchi}, \citenamefont
  {MacDonald},\ and\ \citenamefont {Young}}]{polshyn2020electrical}%
  \BibitemOpen
  \bibfield  {author} {\bibinfo {author} {\bibfnamefont {H.}~\bibnamefont
  {Polshyn}}, \bibinfo {author} {\bibfnamefont {J.}~\bibnamefont {Zhu}},
  \bibinfo {author} {\bibfnamefont {M.~A.}\ \bibnamefont {Kumar}}, \bibinfo
  {author} {\bibfnamefont {Y.}~\bibnamefont {Zhang}}, \bibinfo {author}
  {\bibfnamefont {F.}~\bibnamefont {Yang}}, \bibinfo {author} {\bibfnamefont
  {C.~L.}\ \bibnamefont {Tschirhart}}, \bibinfo {author} {\bibfnamefont
  {M.}~\bibnamefont {Serlin}}, \bibinfo {author} {\bibfnamefont
  {K.}~\bibnamefont {Watanabe}}, \bibinfo {author} {\bibfnamefont
  {T.}~\bibnamefont {Taniguchi}}, \bibinfo {author} {\bibfnamefont {A.~H.}\
  \bibnamefont {MacDonald}},\ and\ \bibinfo {author} {\bibfnamefont {A.~F.}\
  \bibnamefont {Young}},\ }\bibfield  {title} {\bibinfo {title} {Electrical
  switching of magnetic order in an orbital {C}hern insulator},\ }\href
  {https://doi.org/10.1038/s41586-020-2963-8} {\bibfield  {journal} {\bibinfo
  {journal} {Nature}\ }\textbf {\bibinfo {volume} {588}},\ \bibinfo {pages}
  {66} (\bibinfo {year} {2020})}\BibitemShut {NoStop}%
\bibitem [{\citenamefont {Zhu}\ \emph {et~al.}(2020)\citenamefont {Zhu},
  \citenamefont {Su},\ and\ \citenamefont
  {MacDonald}}]{PhysRevLett.125.227702}%
  \BibitemOpen
  \bibfield  {author} {\bibinfo {author} {\bibfnamefont {J.}~\bibnamefont
  {Zhu}}, \bibinfo {author} {\bibfnamefont {J.-J.}\ \bibnamefont {Su}},\ and\
  \bibinfo {author} {\bibfnamefont {A.~H.}\ \bibnamefont {MacDonald}},\
  }\bibfield  {title} {\bibinfo {title} {Voltage-controlled magnetic reversal
  in orbital {C}hern insulators},\ }\href
  {https://doi.org/10.1103/PhysRevLett.125.227702} {\bibfield  {journal}
  {\bibinfo  {journal} {Phys. Rev. Lett.}\ }\textbf {\bibinfo {volume} {125}},\
  \bibinfo {pages} {227702} (\bibinfo {year} {2020})}\BibitemShut {NoStop}%
\bibitem [{\citenamefont {Chen}\ \emph {et~al.}(2021)\citenamefont {Chen},
  \citenamefont {He}, \citenamefont {Zhang}, \citenamefont {Hsieh},
  \citenamefont {Fei}, \citenamefont {Watanabe}, \citenamefont {Taniguchi},
  \citenamefont {Cobden}, \citenamefont {Xu}, \citenamefont {Dean},\ and\
  \citenamefont {Yankowitz}}]{chen2021electrically}%
  \BibitemOpen
  \bibfield  {author} {\bibinfo {author} {\bibfnamefont {S.}~\bibnamefont
  {Chen}}, \bibinfo {author} {\bibfnamefont {M.}~\bibnamefont {He}}, \bibinfo
  {author} {\bibfnamefont {Y.-H.}\ \bibnamefont {Zhang}}, \bibinfo {author}
  {\bibfnamefont {V.}~\bibnamefont {Hsieh}}, \bibinfo {author} {\bibfnamefont
  {Z.}~\bibnamefont {Fei}}, \bibinfo {author} {\bibfnamefont {K.}~\bibnamefont
  {Watanabe}}, \bibinfo {author} {\bibfnamefont {T.}~\bibnamefont {Taniguchi}},
  \bibinfo {author} {\bibfnamefont {D.~H.}\ \bibnamefont {Cobden}}, \bibinfo
  {author} {\bibfnamefont {X.}~\bibnamefont {Xu}}, \bibinfo {author}
  {\bibfnamefont {C.~R.}\ \bibnamefont {Dean}},\ and\ \bibinfo {author}
  {\bibfnamefont {M.}~\bibnamefont {Yankowitz}},\ }\bibfield  {title} {\bibinfo
  {title} {Electrically tunable correlated and topological states in twisted
  monolayer--bilayer graphene},\ }\href
  {https://doi.org/10.1038/s41567-020-01062-6} {\bibfield  {journal} {\bibinfo
  {journal} {Nat. Phys.}\ }\textbf {\bibinfo {volume} {17}},\ \bibinfo {pages}
  {374} (\bibinfo {year} {2021})}\BibitemShut {NoStop}%
\bibitem [{\citenamefont {Tschirhart}\ \emph {et~al.}(2023)\citenamefont
  {Tschirhart}, \citenamefont {Redekop}, \citenamefont {Li}, \citenamefont
  {Li}, \citenamefont {Jiang}, \citenamefont {Arp}, \citenamefont {Sheekey},
  \citenamefont {Taniguchi}, \citenamefont {Watanabe}, \citenamefont {Huber},
  \citenamefont {Mak}, \citenamefont {Shan},\ and\ \citenamefont
  {Young}}]{tschirhart2023intrinsic}%
  \BibitemOpen
  \bibfield  {author} {\bibinfo {author} {\bibfnamefont {C.~L.}\ \bibnamefont
  {Tschirhart}}, \bibinfo {author} {\bibfnamefont {E.}~\bibnamefont {Redekop}},
  \bibinfo {author} {\bibfnamefont {L.}~\bibnamefont {Li}}, \bibinfo {author}
  {\bibfnamefont {T.}~\bibnamefont {Li}}, \bibinfo {author} {\bibfnamefont
  {S.}~\bibnamefont {Jiang}}, \bibinfo {author} {\bibfnamefont
  {T.}~\bibnamefont {Arp}}, \bibinfo {author} {\bibfnamefont {O.}~\bibnamefont
  {Sheekey}}, \bibinfo {author} {\bibfnamefont {T.}~\bibnamefont {Taniguchi}},
  \bibinfo {author} {\bibfnamefont {K.}~\bibnamefont {Watanabe}}, \bibinfo
  {author} {\bibfnamefont {M.~E.}\ \bibnamefont {Huber}}, \bibinfo {author}
  {\bibfnamefont {K.~F.}\ \bibnamefont {Mak}}, \bibinfo {author} {\bibfnamefont
  {J.}~\bibnamefont {Shan}},\ and\ \bibinfo {author} {\bibfnamefont {A.~F.}\
  \bibnamefont {Young}},\ }\bibfield  {title} {\bibinfo {title} {Intrinsic spin
  {H}all torque in a moir{\'e} {C}hern magnet},\ }\href
  {https://doi.org/10.1038/s41567-023-01979-8} {\bibfield  {journal} {\bibinfo
  {journal} {Nat. Phys.}\ }\textbf {\bibinfo {volume} {19}},\ \bibinfo {pages}
  {807} (\bibinfo {year} {2023})}\BibitemShut {NoStop}%
\bibitem [{\citenamefont {Wu}(2017)}]{Wu2017RareEarthQAH}%
  \BibitemOpen
  \bibfield  {author} {\bibinfo {author} {\bibfnamefont {M.}~\bibnamefont
  {Wu}},\ }\bibfield  {title} {\bibinfo {title} {High-temperature intrinsic
  quantum anomalous {H}all effect in rare earth monohalide},\ }\href
  {https://doi.org/10.1088/2053-1583/aa5c63} {\bibfield  {journal} {\bibinfo
  {journal} {2D Mater.}\ }\textbf {\bibinfo {volume} {4}},\ \bibinfo {pages}
  {021014} (\bibinfo {year} {2017})}\BibitemShut {NoStop}%
\bibitem [{\citenamefont {Nie}\ \emph {et~al.}(2019)\citenamefont {Nie},
  \citenamefont {Weng},\ and\ \citenamefont {Prinz}}]{PhysRevB.99.035125}%
  \BibitemOpen
  \bibfield  {author} {\bibinfo {author} {\bibfnamefont {S.}~\bibnamefont
  {Nie}}, \bibinfo {author} {\bibfnamefont {H.}~\bibnamefont {Weng}},\ and\
  \bibinfo {author} {\bibfnamefont {F.~B.}\ \bibnamefont {Prinz}},\ }\bibfield
  {title} {\bibinfo {title} {Topological nodal-line semimetals in ferromagnetic
  rare-earth-metal monohalides},\ }\href
  {https://doi.org/10.1103/PhysRevB.99.035125} {\bibfield  {journal} {\bibinfo
  {journal} {Phys. Rev. B}\ }\textbf {\bibinfo {volume} {99}},\ \bibinfo
  {pages} {035125} (\bibinfo {year} {2019})}\BibitemShut {NoStop}%
\bibitem [{\citenamefont {Guo}\ \emph {et~al.}(2020)\citenamefont {Guo},
  \citenamefont {Liu}, \citenamefont {Liu}, \citenamefont {Li},\ and\
  \citenamefont {Wang}}]{Pr2QAHE}%
  \BibitemOpen
  \bibfield  {author} {\bibinfo {author} {\bibfnamefont {X.}~\bibnamefont
  {Guo}}, \bibinfo {author} {\bibfnamefont {Z.}~\bibnamefont {Liu}}, \bibinfo
  {author} {\bibfnamefont {B.}~\bibnamefont {Liu}}, \bibinfo {author}
  {\bibfnamefont {Q.}~\bibnamefont {Li}},\ and\ \bibinfo {author}
  {\bibfnamefont {Z.}~\bibnamefont {Wang}},\ }\bibfield  {title} {\bibinfo
  {title} {Non-collinear orbital-induced planar quantum anomalous {H}all
  effect},\ }\href {https://doi.org/10.1021/acs.nanolett.0c03136} {\bibfield
  {journal} {\bibinfo  {journal} {Nano Lett.}\ }\textbf {\bibinfo {volume}
  {20}},\ \bibinfo {pages} {7606} (\bibinfo {year} {2020})}\BibitemShut
  {NoStop}%
\bibitem [{\citenamefont {Guo}\ and\ \citenamefont
  {Franz}(2009)}]{PhysRevB.80.113102}%
  \BibitemOpen
  \bibfield  {author} {\bibinfo {author} {\bibfnamefont {H.-M.}\ \bibnamefont
  {Guo}}\ and\ \bibinfo {author} {\bibfnamefont {M.}~\bibnamefont {Franz}},\
  }\bibfield  {title} {\bibinfo {title} {Topological insulator on the kagome
  lattice},\ }\href {https://doi.org/10.1103/PhysRevB.80.113102} {\bibfield
  {journal} {\bibinfo  {journal} {Phys. Rev. B}\ }\textbf {\bibinfo {volume}
  {80}},\ \bibinfo {pages} {113102} (\bibinfo {year} {2009})}\BibitemShut
  {NoStop}%
\bibitem [{\citenamefont {Ren}\ \emph {et~al.}(2018)\citenamefont {Ren},
  \citenamefont {Zeng}, \citenamefont {Zhu},\ and\ \citenamefont
  {Sheng}}]{PhysRevB.98.205146}%
  \BibitemOpen
  \bibfield  {author} {\bibinfo {author} {\bibfnamefont {Y.}~\bibnamefont
  {Ren}}, \bibinfo {author} {\bibfnamefont {T.-S.}\ \bibnamefont {Zeng}},
  \bibinfo {author} {\bibfnamefont {W.}~\bibnamefont {Zhu}},\ and\ \bibinfo
  {author} {\bibfnamefont {D.~N.}\ \bibnamefont {Sheng}},\ }\bibfield  {title}
  {\bibinfo {title} {Quantum anomalous {H}all phase stabilized via realistic
  interactions on a kagome lattice},\ }\href
  {https://doi.org/10.1103/PhysRevB.98.205146} {\bibfield  {journal} {\bibinfo
  {journal} {Phys. Rev. B}\ }\textbf {\bibinfo {volume} {98}},\ \bibinfo
  {pages} {205146} (\bibinfo {year} {2018})}\BibitemShut {NoStop}%
\bibitem [{\citenamefont {Zhang}\ \emph {et~al.}(2021)\citenamefont {Zhang},
  \citenamefont {You}, \citenamefont {Ma}, \citenamefont {Gu},\ and\
  \citenamefont {Su}}]{PhysRevB.103.014410}%
  \BibitemOpen
  \bibfield  {author} {\bibinfo {author} {\bibfnamefont {Z.}~\bibnamefont
  {Zhang}}, \bibinfo {author} {\bibfnamefont {J.-Y.}\ \bibnamefont {You}},
  \bibinfo {author} {\bibfnamefont {X.-Y.}\ \bibnamefont {Ma}}, \bibinfo
  {author} {\bibfnamefont {B.}~\bibnamefont {Gu}},\ and\ \bibinfo {author}
  {\bibfnamefont {G.}~\bibnamefont {Su}},\ }\bibfield  {title} {\bibinfo
  {title} {Kagome quantum anomalous {H}all effect with high {C}hern number and
  large band gap},\ }\href {https://doi.org/10.1103/PhysRevB.103.014410}
  {\bibfield  {journal} {\bibinfo  {journal} {Phys. Rev. B}\ }\textbf {\bibinfo
  {volume} {103}},\ \bibinfo {pages} {014410} (\bibinfo {year}
  {2021})}\BibitemShut {NoStop}%
\bibitem [{\citenamefont {Wang}\ \emph {et~al.}(2024)\citenamefont {Wang},
  \citenamefont {Lei}, \citenamefont {Qi},\ and\ \citenamefont
  {Felser}}]{Wang2024TopologicalQM}%
  \BibitemOpen
  \bibfield  {author} {\bibinfo {author} {\bibfnamefont {Q.}~\bibnamefont
  {Wang}}, \bibinfo {author} {\bibfnamefont {H.}~\bibnamefont {Lei}}, \bibinfo
  {author} {\bibfnamefont {Y.}~\bibnamefont {Qi}},\ and\ \bibinfo {author}
  {\bibfnamefont {C.}~\bibnamefont {Felser}},\ }\bibfield  {title} {\bibinfo
  {title} {Topological quantum materials with kagome lattice},\ }\href
  {https://doi.org/10.1021/accountsmr.3c00291} {\bibfield  {journal} {\bibinfo
  {journal} {Acc. Mater. Res.}\ }\textbf {\bibinfo {volume} {5}},\ \bibinfo
  {pages} {786} (\bibinfo {year} {2024})}\BibitemShut {NoStop}%
\bibitem [{\citenamefont {Liu}\ \emph {et~al.}(2013)\citenamefont {Liu},
  \citenamefont {Wang}, \citenamefont {Mei}, \citenamefont {Wu},\ and\
  \citenamefont {Liu}}]{PhysRevLett.110.106804}%
  \BibitemOpen
  \bibfield  {author} {\bibinfo {author} {\bibfnamefont {Z.}~\bibnamefont
  {Liu}}, \bibinfo {author} {\bibfnamefont {Z.-F.}\ \bibnamefont {Wang}},
  \bibinfo {author} {\bibfnamefont {J.-W.}\ \bibnamefont {Mei}}, \bibinfo
  {author} {\bibfnamefont {Y.-S.}\ \bibnamefont {Wu}},\ and\ \bibinfo {author}
  {\bibfnamefont {F.}~\bibnamefont {Liu}},\ }\bibfield  {title} {\bibinfo
  {title} {Flat {C}hern band in a two-dimensional organometallic framework},\
  }\href {https://doi.org/10.1103/PhysRevLett.110.106804} {\bibfield  {journal}
  {\bibinfo  {journal} {Phys. Rev. Lett.}\ }\textbf {\bibinfo {volume} {110}},\
  \bibinfo {pages} {106804} (\bibinfo {year} {2013})}\BibitemShut {NoStop}%
\bibitem [{\citenamefont {Wang}\ \emph
  {et~al.}(2013{\natexlab{a}})\citenamefont {Wang}, \citenamefont {Liu},\ and\
  \citenamefont {Liu}}]{PhysRevLett.110.196801}%
  \BibitemOpen
  \bibfield  {author} {\bibinfo {author} {\bibfnamefont {Z.~F.}\ \bibnamefont
  {Wang}}, \bibinfo {author} {\bibfnamefont {Z.}~\bibnamefont {Liu}},\ and\
  \bibinfo {author} {\bibfnamefont {F.}~\bibnamefont {Liu}},\ }\bibfield
  {title} {\bibinfo {title} {Quantum anomalous {H}all effect in 2{D} organic
  topological insulators},\ }\href
  {https://doi.org/10.1103/PhysRevLett.110.196801} {\bibfield  {journal}
  {\bibinfo  {journal} {Phys. Rev. Lett.}\ }\textbf {\bibinfo {volume} {110}},\
  \bibinfo {pages} {196801} (\bibinfo {year} {2013}{\natexlab{a}})}\BibitemShut
  {NoStop}%
\bibitem [{\citenamefont {Makhfudz}\ \emph {et~al.}(2024)\citenamefont
  {Makhfudz}, \citenamefont {Cherkasskii}, \citenamefont {Alipourzadeh},
  \citenamefont {Hajati}, \citenamefont {Lombardo}, \citenamefont
  {Sch\"{a}fer}, \citenamefont {Viola~Kusminskiy},\ and\ \citenamefont
  {Hayn}}]{PhysRevB.110.235130}%
  \BibitemOpen
  \bibfield  {author} {\bibinfo {author} {\bibfnamefont {I.}~\bibnamefont
  {Makhfudz}}, \bibinfo {author} {\bibfnamefont {M.}~\bibnamefont
  {Cherkasskii}}, \bibinfo {author} {\bibfnamefont {M.}~\bibnamefont
  {Alipourzadeh}}, \bibinfo {author} {\bibfnamefont {Y.}~\bibnamefont
  {Hajati}}, \bibinfo {author} {\bibfnamefont {P.}~\bibnamefont {Lombardo}},
  \bibinfo {author} {\bibfnamefont {S.}~\bibnamefont {Sch\"{a}fer}}, \bibinfo
  {author} {\bibfnamefont {S.}~\bibnamefont {Viola~Kusminskiy}},\ and\ \bibinfo
  {author} {\bibfnamefont {R.}~\bibnamefont {Hayn}},\ }\bibfield  {title}
  {\bibinfo {title} {Quantum anomalous {H}all effect in $d$-electron kagome
  systems: {C}hern insulating states from transverse spin-orbit coupling},\
  }\href {https://doi.org/10.1103/PhysRevB.110.235130} {\bibfield  {journal}
  {\bibinfo  {journal} {Phys. Rev. B}\ }\textbf {\bibinfo {volume} {110}},\
  \bibinfo {pages} {235130} (\bibinfo {year} {2024})}\BibitemShut {NoStop}%
\bibitem [{\citenamefont {Guo}\ \emph {et~al.}(2026)\citenamefont {Guo},
  \citenamefont {Nie},\ and\ \citenamefont {Prinz}}]{YbKagome2026}%
  \BibitemOpen
  \bibfield  {author} {\bibinfo {author} {\bibfnamefont {J.}~\bibnamefont
  {Guo}}, \bibinfo {author} {\bibfnamefont {S.}~\bibnamefont {Nie}},\ and\
  \bibinfo {author} {\bibfnamefont {F.~B.}\ \bibnamefont {Prinz}},\ }\bibfield
  {title} {\bibinfo {title} {Layer-dependent and gate-tunable {C}hern numbers
  in 2{D} kagome ferromagnet {Yb$_2$(C$_6$H$_4$)$_3$} with a large band gap},\
  }\href {https://doi.org/10.1038/s41524-026-01991-5} {\bibfield  {journal}
  {\bibinfo  {journal} {npj Comput. Mater.}\ }\textbf {\bibinfo {volume}
  {12}},\ \bibinfo {pages} {111} (\bibinfo {year} {2026})}\BibitemShut
  {NoStop}%
\bibitem [{\citenamefont {Kresse}\ and\ \citenamefont
  {Hafner}(1993)}]{PhysRevB.47.558}%
  \BibitemOpen
  \bibfield  {author} {\bibinfo {author} {\bibfnamefont {G.}~\bibnamefont
  {Kresse}}\ and\ \bibinfo {author} {\bibfnamefont {J.}~\bibnamefont
  {Hafner}},\ }\bibfield  {title} {\bibinfo {title} {Ab initio molecular
  dynamics for liquid metals},\ }\href
  {https://doi.org/10.1103/PhysRevB.47.558} {\bibfield  {journal} {\bibinfo
  {journal} {Phys. Rev. B}\ }\textbf {\bibinfo {volume} {47}},\ \bibinfo
  {pages} {558} (\bibinfo {year} {1993})}\BibitemShut {NoStop}%
\bibitem [{\citenamefont {Kresse}\ and\ \citenamefont
  {Furthm\"uller}(1996)}]{PhysRevB.54.11169}%
  \BibitemOpen
  \bibfield  {author} {\bibinfo {author} {\bibfnamefont {G.}~\bibnamefont
  {Kresse}}\ and\ \bibinfo {author} {\bibfnamefont {J.}~\bibnamefont
  {Furthm\"uller}},\ }\bibfield  {title} {\bibinfo {title} {Efficient iterative
  schemes for ab initio total-energy calculations using a plane-wave basis
  set},\ }\href {https://doi.org/10.1103/PhysRevB.54.11169} {\bibfield
  {journal} {\bibinfo  {journal} {Phys. Rev. B}\ }\textbf {\bibinfo {volume}
  {54}},\ \bibinfo {pages} {11169} (\bibinfo {year} {1996})}\BibitemShut
  {NoStop}%
\bibitem [{\citenamefont {Bl\"ochl}(1994)}]{PhysRevB.50.17953}%
  \BibitemOpen
  \bibfield  {author} {\bibinfo {author} {\bibfnamefont {P.~E.}\ \bibnamefont
  {Bl\"ochl}},\ }\bibfield  {title} {\bibinfo {title} {Projector augmented-wave
  method},\ }\href {https://doi.org/10.1103/PhysRevB.50.17953} {\bibfield
  {journal} {\bibinfo  {journal} {Phys. Rev. B}\ }\textbf {\bibinfo {volume}
  {50}},\ \bibinfo {pages} {17953} (\bibinfo {year} {1994})}\BibitemShut
  {NoStop}%
\bibitem [{\citenamefont {Perdew}\ \emph {et~al.}(1996)\citenamefont {Perdew},
  \citenamefont {Burke},\ and\ \citenamefont
  {Ernzerhof}}]{PhysRevLett.77.3865}%
  \BibitemOpen
  \bibfield  {author} {\bibinfo {author} {\bibfnamefont {J.~P.}\ \bibnamefont
  {Perdew}}, \bibinfo {author} {\bibfnamefont {K.}~\bibnamefont {Burke}},\ and\
  \bibinfo {author} {\bibfnamefont {M.}~\bibnamefont {Ernzerhof}},\ }\bibfield
  {title} {\bibinfo {title} {Generalized gradient approximation made simple},\
  }\href {https://doi.org/10.1103/PhysRevLett.77.3865} {\bibfield  {journal}
  {\bibinfo  {journal} {Phys. Rev. Lett.}\ }\textbf {\bibinfo {volume} {77}},\
  \bibinfo {pages} {3865} (\bibinfo {year} {1996})}\BibitemShut {NoStop}%
\bibitem [{\citenamefont {Dudarev}\ \emph {et~al.}(1998)\citenamefont
  {Dudarev}, \citenamefont {Botton}, \citenamefont {Savrasov}, \citenamefont
  {Humphreys},\ and\ \citenamefont {Sutton}}]{PhysRevB.57.1505}%
  \BibitemOpen
  \bibfield  {author} {\bibinfo {author} {\bibfnamefont {S.~L.}\ \bibnamefont
  {Dudarev}}, \bibinfo {author} {\bibfnamefont {G.~A.}\ \bibnamefont {Botton}},
  \bibinfo {author} {\bibfnamefont {S.~Y.}\ \bibnamefont {Savrasov}}, \bibinfo
  {author} {\bibfnamefont {C.~J.}\ \bibnamefont {Humphreys}},\ and\ \bibinfo
  {author} {\bibfnamefont {A.~P.}\ \bibnamefont {Sutton}},\ }\bibfield  {title}
  {\bibinfo {title} {Electron-energy-loss spectra and the structural stability
  of nickel oxide: An {LSDA}+{U} study},\ }\href
  {https://doi.org/10.1103/PhysRevB.57.1505} {\bibfield  {journal} {\bibinfo
  {journal} {Phys. Rev. B}\ }\textbf {\bibinfo {volume} {57}},\ \bibinfo
  {pages} {1505} (\bibinfo {year} {1998})}\BibitemShut {NoStop}%
\bibitem [{\citenamefont {Grimme}\ \emph {et~al.}(2011)\citenamefont {Grimme},
  \citenamefont {Ehrlich},\ and\ \citenamefont {Goerigk}}]{Grimme2011EffectOT}%
  \BibitemOpen
  \bibfield  {author} {\bibinfo {author} {\bibfnamefont {S.}~\bibnamefont
  {Grimme}}, \bibinfo {author} {\bibfnamefont {S.}~\bibnamefont {Ehrlich}},\
  and\ \bibinfo {author} {\bibfnamefont {L.}~\bibnamefont {Goerigk}},\
  }\bibfield  {title} {\bibinfo {title} {Effect of the damping function in
  dispersion corrected density functional theory},\ }\href
  {https://doi.org/10.1002/jcc.21759} {\bibfield  {journal} {\bibinfo
  {journal} {J. Comput. Chem.}\ }\textbf {\bibinfo {volume} {32}},\ \bibinfo
  {pages} {1456} (\bibinfo {year} {2011})}\BibitemShut {NoStop}%
\bibitem [{\citenamefont {Monkhorst}\ and\ \citenamefont
  {Pack}(1976)}]{PhysRevB.13.5188}%
  \BibitemOpen
  \bibfield  {author} {\bibinfo {author} {\bibfnamefont {H.~J.}\ \bibnamefont
  {Monkhorst}}\ and\ \bibinfo {author} {\bibfnamefont {J.~D.}\ \bibnamefont
  {Pack}},\ }\bibfield  {title} {\bibinfo {title} {Special points for
  {B}rillouin-zone integrations},\ }\href
  {https://doi.org/10.1103/PhysRevB.13.5188} {\bibfield  {journal} {\bibinfo
  {journal} {Phys. Rev. B}\ }\textbf {\bibinfo {volume} {13}},\ \bibinfo
  {pages} {5188} (\bibinfo {year} {1976})}\BibitemShut {NoStop}%
\bibitem [{\citenamefont {Togo}\ and\ \citenamefont
  {Tanaka}(2015)}]{TOGO20151}%
  \BibitemOpen
  \bibfield  {author} {\bibinfo {author} {\bibfnamefont {A.}~\bibnamefont
  {Togo}}\ and\ \bibinfo {author} {\bibfnamefont {I.}~\bibnamefont {Tanaka}},\
  }\bibfield  {title} {\bibinfo {title} {First principles phonon calculations
  in materials science},\ }\href
  {https://doi.org/10.1016/j.scriptamat.2015.07.021} {\bibfield  {journal}
  {\bibinfo  {journal} {Scr. Mater.}\ }\textbf {\bibinfo {volume} {108}},\
  \bibinfo {pages} {1} (\bibinfo {year} {2015})}\BibitemShut {NoStop}%
\bibitem [{\citenamefont {Nos{\'e}}(1984)}]{nose1984unified}%
  \BibitemOpen
  \bibfield  {author} {\bibinfo {author} {\bibfnamefont {S.}~\bibnamefont
  {Nos{\'e}}},\ }\bibfield  {title} {\bibinfo {title} {A unified formulation of
  the constant temperature molecular dynamics methods},\ }\href
  {https://doi.org/10.1063/1.447334} {\bibfield  {journal} {\bibinfo  {journal}
  {J. Chem. Phys.}\ }\textbf {\bibinfo {volume} {81}},\ \bibinfo {pages} {511}
  (\bibinfo {year} {1984})}\BibitemShut {NoStop}%
\bibitem [{\citenamefont {Hoover}(1985)}]{hoover1985canonical}%
  \BibitemOpen
  \bibfield  {author} {\bibinfo {author} {\bibfnamefont {W.~G.}\ \bibnamefont
  {Hoover}},\ }\bibfield  {title} {\bibinfo {title} {Canonical dynamics:
  Equilibrium phase-space distributions},\ }\href
  {https://doi.org/10.1103/PhysRevA.31.1695} {\bibfield  {journal} {\bibinfo
  {journal} {Phys. Rev. A}\ }\textbf {\bibinfo {volume} {31}},\ \bibinfo
  {pages} {1695} (\bibinfo {year} {1985})}\BibitemShut {NoStop}%
\bibitem [{\citenamefont {Liechtenstein}\ \emph {et~al.}(1987)\citenamefont
  {Liechtenstein}, \citenamefont {Katsnelson}, \citenamefont {Antropov},\ and\
  \citenamefont {Gubanov}}]{liechtenstein1987local}%
  \BibitemOpen
  \bibfield  {author} {\bibinfo {author} {\bibfnamefont {A.~I.}\ \bibnamefont
  {Liechtenstein}}, \bibinfo {author} {\bibfnamefont {M.}~\bibnamefont
  {Katsnelson}}, \bibinfo {author} {\bibfnamefont {V.}~\bibnamefont
  {Antropov}},\ and\ \bibinfo {author} {\bibfnamefont {V.}~\bibnamefont
  {Gubanov}},\ }\bibfield  {title} {\bibinfo {title} {Local spin density
  functional approach to the theory of exchange interactions in ferromagnetic
  metals and alloys},\ }\href {https://doi.org/10.1016/0304-8853(87)90721-9}
  {\bibfield  {journal} {\bibinfo  {journal} {J. Magn. Magn. Mater.}\ }\textbf
  {\bibinfo {volume} {67}},\ \bibinfo {pages} {65} (\bibinfo {year}
  {1987})}\BibitemShut {NoStop}%
\bibitem [{\citenamefont {Terasawa}\ \emph {et~al.}(2019)\citenamefont
  {Terasawa}, \citenamefont {Matsumoto}, \citenamefont {Ozaki},\ and\
  \citenamefont {Gohda}}]{terasawa2019efficient}%
  \BibitemOpen
  \bibfield  {author} {\bibinfo {author} {\bibfnamefont {A.}~\bibnamefont
  {Terasawa}}, \bibinfo {author} {\bibfnamefont {M.}~\bibnamefont {Matsumoto}},
  \bibinfo {author} {\bibfnamefont {T.}~\bibnamefont {Ozaki}},\ and\ \bibinfo
  {author} {\bibfnamefont {Y.}~\bibnamefont {Gohda}},\ }\bibfield  {title}
  {\bibinfo {title} {Efficient algorithm based on liechtenstein method for
  computing exchange coupling constants using localized basis set},\ }\href
  {https://doi.org/10.7566/JPSJ.88.114706} {\bibfield  {journal} {\bibinfo
  {journal} {J. Phys. Soc. Jpn.}\ }\textbf {\bibinfo {volume} {88}},\ \bibinfo
  {pages} {114706} (\bibinfo {year} {2019})}\BibitemShut {NoStop}%
\bibitem [{\citenamefont {Pajda}\ \emph {et~al.}(2001)\citenamefont {Pajda},
  \citenamefont {Kudrnovsk\'y}, \citenamefont {Turek}, \citenamefont {Drchal},\
  and\ \citenamefont {Bruno}}]{PhysRevB.64.174402}%
  \BibitemOpen
  \bibfield  {author} {\bibinfo {author} {\bibfnamefont {M.}~\bibnamefont
  {Pajda}}, \bibinfo {author} {\bibfnamefont {J.}~\bibnamefont {Kudrnovsk\'y}},
  \bibinfo {author} {\bibfnamefont {I.}~\bibnamefont {Turek}}, \bibinfo
  {author} {\bibfnamefont {V.}~\bibnamefont {Drchal}},\ and\ \bibinfo {author}
  {\bibfnamefont {P.}~\bibnamefont {Bruno}},\ }\bibfield  {title} {\bibinfo
  {title} {Ab initio calculations of exchange interactions, spin-wave stiffness
  constants, and {C}urie temperatures of {Fe}, {Co}, and {Ni}},\ }\href
  {https://doi.org/10.1103/PhysRevB.64.174402} {\bibfield  {journal} {\bibinfo
  {journal} {Phys. Rev. B}\ }\textbf {\bibinfo {volume} {64}},\ \bibinfo
  {pages} {174402} (\bibinfo {year} {2001})}\BibitemShut {NoStop}%
\bibitem [{\citenamefont {Smidstrup}\ \emph {et~al.}(2020)\citenamefont
  {Smidstrup}, \citenamefont {Markussen}, \citenamefont {Vancraeyveld},
  \citenamefont {Wellendorff}, \citenamefont {Schneider}, \citenamefont
  {Gunst}, \citenamefont {Verstichel}, \citenamefont {Stradi}, \citenamefont
  {Khomyakov}, \citenamefont {Vej-Hansen}, \citenamefont {Lee}, \citenamefont
  {Chill}, \citenamefont {Rasmussen}, \citenamefont {Penazzi}, \citenamefont
  {Corsetti}, \citenamefont {Ojanper{\"a}}, \citenamefont {Jensen},
  \citenamefont {Palsgaard}, \citenamefont {Martinez}, \citenamefont {Blom},
  \citenamefont {Brandbyge},\ and\ \citenamefont
  {Stokbro}}]{smidstrup2019quantumatk}%
  \BibitemOpen
  \bibfield  {author} {\bibinfo {author} {\bibfnamefont {S.}~\bibnamefont
  {Smidstrup}}, \bibinfo {author} {\bibfnamefont {T.}~\bibnamefont
  {Markussen}}, \bibinfo {author} {\bibfnamefont {P.}~\bibnamefont
  {Vancraeyveld}}, \bibinfo {author} {\bibfnamefont {J.}~\bibnamefont
  {Wellendorff}}, \bibinfo {author} {\bibfnamefont {J.}~\bibnamefont
  {Schneider}}, \bibinfo {author} {\bibfnamefont {T.}~\bibnamefont {Gunst}},
  \bibinfo {author} {\bibfnamefont {B.}~\bibnamefont {Verstichel}}, \bibinfo
  {author} {\bibfnamefont {D.}~\bibnamefont {Stradi}}, \bibinfo {author}
  {\bibfnamefont {P.~A.}\ \bibnamefont {Khomyakov}}, \bibinfo {author}
  {\bibfnamefont {U.~G.}\ \bibnamefont {Vej-Hansen}}, \bibinfo {author}
  {\bibfnamefont {M.-E.}\ \bibnamefont {Lee}}, \bibinfo {author} {\bibfnamefont
  {S.~T.}\ \bibnamefont {Chill}}, \bibinfo {author} {\bibfnamefont
  {F.}~\bibnamefont {Rasmussen}}, \bibinfo {author} {\bibfnamefont
  {G.}~\bibnamefont {Penazzi}}, \bibinfo {author} {\bibfnamefont
  {F.}~\bibnamefont {Corsetti}}, \bibinfo {author} {\bibfnamefont
  {A.}~\bibnamefont {Ojanper{\"a}}}, \bibinfo {author} {\bibfnamefont
  {K.}~\bibnamefont {Jensen}}, \bibinfo {author} {\bibfnamefont {M.~L.~N.}\
  \bibnamefont {Palsgaard}}, \bibinfo {author} {\bibfnamefont {U.}~\bibnamefont
  {Martinez}}, \bibinfo {author} {\bibfnamefont {A.}~\bibnamefont {Blom}},
  \bibinfo {author} {\bibfnamefont {M.}~\bibnamefont {Brandbyge}},\ and\
  \bibinfo {author} {\bibfnamefont {K.}~\bibnamefont {Stokbro}},\ }\bibfield
  {title} {\bibinfo {title} {{QuantumATK}: an integrated platform of electronic
  and atomic-scale modelling tools},\ }\href
  {https://doi.org/10.1088/1361-648X/ab4007} {\bibfield  {journal} {\bibinfo
  {journal} {J. Phys.: Condens. Matter}\ }\textbf {\bibinfo {volume} {32}},\
  \bibinfo {pages} {015901} (\bibinfo {year} {2020})}\BibitemShut {NoStop}%
\bibitem [{\citenamefont {Evans}\ \emph {et~al.}(2014)\citenamefont {Evans},
  \citenamefont {Fan}, \citenamefont {Chureemart}, \citenamefont {Ostler},
  \citenamefont {Ellis},\ and\ \citenamefont {Chantrell}}]{evans2014atomistic}%
  \BibitemOpen
  \bibfield  {author} {\bibinfo {author} {\bibfnamefont {R.~F.}\ \bibnamefont
  {Evans}}, \bibinfo {author} {\bibfnamefont {W.~J.}\ \bibnamefont {Fan}},
  \bibinfo {author} {\bibfnamefont {P.}~\bibnamefont {Chureemart}}, \bibinfo
  {author} {\bibfnamefont {T.~A.}\ \bibnamefont {Ostler}}, \bibinfo {author}
  {\bibfnamefont {M.~O.}\ \bibnamefont {Ellis}},\ and\ \bibinfo {author}
  {\bibfnamefont {R.~W.}\ \bibnamefont {Chantrell}},\ }\bibfield  {title}
  {\bibinfo {title} {Atomistic spin model simulations of magnetic
  nanomaterials},\ }\href {https://doi.org/10.1088/0953-8984/26/10/103202}
  {\bibfield  {journal} {\bibinfo  {journal} {J. Phys.: Condens. Matter}\
  }\textbf {\bibinfo {volume} {26}},\ \bibinfo {pages} {103202} (\bibinfo
  {year} {2014})}\BibitemShut {NoStop}%
\bibitem [{\citenamefont {Mostofi}\ \emph {et~al.}(2014)\citenamefont
  {Mostofi}, \citenamefont {Yates}, \citenamefont {Pizzi}, \citenamefont {Lee},
  \citenamefont {Souza}, \citenamefont {Vanderbilt},\ and\ \citenamefont
  {Marzari}}]{Mostofi2014AnUV}%
  \BibitemOpen
  \bibfield  {author} {\bibinfo {author} {\bibfnamefont {A.~A.}\ \bibnamefont
  {Mostofi}}, \bibinfo {author} {\bibfnamefont {J.~R.}\ \bibnamefont {Yates}},
  \bibinfo {author} {\bibfnamefont {G.}~\bibnamefont {Pizzi}}, \bibinfo
  {author} {\bibfnamefont {Y.-S.}\ \bibnamefont {Lee}}, \bibinfo {author}
  {\bibfnamefont {I.}~\bibnamefont {Souza}}, \bibinfo {author} {\bibfnamefont
  {D.}~\bibnamefont {Vanderbilt}},\ and\ \bibinfo {author} {\bibfnamefont
  {N.}~\bibnamefont {Marzari}},\ }\bibfield  {title} {\bibinfo {title} {An
  updated version of wannier90: A tool for obtaining maximally-localised
  {W}annier functions},\ }\href {https://doi.org/10.1016/j.cpc.2014.05.003}
  {\bibfield  {journal} {\bibinfo  {journal} {Comput. Phys. Commun.}\ }\textbf
  {\bibinfo {volume} {185}},\ \bibinfo {pages} {2309} (\bibinfo {year}
  {2014})}\BibitemShut {NoStop}%
\bibitem [{\citenamefont {Sancho}\ \emph {et~al.}(1985)\citenamefont {Sancho},
  \citenamefont {Sancho}, \citenamefont {Sancho},\ and\ \citenamefont
  {Rubio}}]{Sancho1985HighlyCS}%
  \BibitemOpen
  \bibfield  {author} {\bibinfo {author} {\bibfnamefont {M.~P.~L.}\
  \bibnamefont {Sancho}}, \bibinfo {author} {\bibfnamefont {J.~M.~L.}\
  \bibnamefont {Sancho}}, \bibinfo {author} {\bibfnamefont {J.~M.~L.}\
  \bibnamefont {Sancho}},\ and\ \bibinfo {author} {\bibfnamefont
  {J.}~\bibnamefont {Rubio}},\ }\bibfield  {title} {\bibinfo {title} {Highly
  convergent schemes for the calculation of bulk and surface {G}reen
  functions},\ }\href {https://doi.org/10.1088/0305-4608/15/4/009} {\bibfield
  {journal} {\bibinfo  {journal} {J. Phys. F: Met. Phys.}\ }\textbf {\bibinfo
  {volume} {15}},\ \bibinfo {pages} {851} (\bibinfo {year} {1985})}\BibitemShut
  {NoStop}%
\bibitem [{\citenamefont {Wu}\ \emph {et~al.}(2018)\citenamefont {Wu},
  \citenamefont {Zhang}, \citenamefont {Song}, \citenamefont {Troyer},\ and\
  \citenamefont {Soluyanov}}]{Wu2017WannierToolsAO}%
  \BibitemOpen
  \bibfield  {author} {\bibinfo {author} {\bibfnamefont {Q.}~\bibnamefont
  {Wu}}, \bibinfo {author} {\bibfnamefont {S.}~\bibnamefont {Zhang}}, \bibinfo
  {author} {\bibfnamefont {H.-F.}\ \bibnamefont {Song}}, \bibinfo {author}
  {\bibfnamefont {M.}~\bibnamefont {Troyer}},\ and\ \bibinfo {author}
  {\bibfnamefont {A.~A.}\ \bibnamefont {Soluyanov}},\ }\bibfield  {title}
  {\bibinfo {title} {{WannierTools}: An open-source software package for novel
  topological materials},\ }\href {https://doi.org/10.1016/j.cpc.2017.09.033}
  {\bibfield  {journal} {\bibinfo  {journal} {Comput. Phys. Commun.}\ }\textbf
  {\bibinfo {volume} {224}},\ \bibinfo {pages} {405} (\bibinfo {year}
  {2018})}\BibitemShut {NoStop}%
\bibitem [{\citenamefont {Fukui}\ \emph {et~al.}(2005)\citenamefont {Fukui},
  \citenamefont {Hatsugai},\ and\ \citenamefont {Suzuki}}]{FHS2005}%
  \BibitemOpen
  \bibfield  {author} {\bibinfo {author} {\bibfnamefont {T.}~\bibnamefont
  {Fukui}}, \bibinfo {author} {\bibfnamefont {Y.}~\bibnamefont {Hatsugai}},\
  and\ \bibinfo {author} {\bibfnamefont {H.}~\bibnamefont {Suzuki}},\
  }\bibfield  {title} {\bibinfo {title} {{C}hern numbers in discretized
  {B}rillouin zone: Efficient method of computing (spin) {H}all conductances},\
  }\href {https://doi.org/10.1143/JPSJ.74.1674} {\bibfield  {journal} {\bibinfo
   {journal} {J. Phys. Soc. Jpn.}\ }\textbf {\bibinfo {volume} {74}},\ \bibinfo
  {pages} {1674} (\bibinfo {year} {2005})}\BibitemShut {NoStop}%
\bibitem [{\citenamefont {Wei}\ \emph {et~al.}(2009)\citenamefont {Wei},
  \citenamefont {Fragneaud}, \citenamefont {Marianetti},\ and\ \citenamefont
  {Kysar}}]{PhysRevB.80.205407}%
  \BibitemOpen
  \bibfield  {author} {\bibinfo {author} {\bibfnamefont {X.}~\bibnamefont
  {Wei}}, \bibinfo {author} {\bibfnamefont {B.}~\bibnamefont {Fragneaud}},
  \bibinfo {author} {\bibfnamefont {C.~A.}\ \bibnamefont {Marianetti}},\ and\
  \bibinfo {author} {\bibfnamefont {J.~W.}\ \bibnamefont {Kysar}},\ }\bibfield
  {title} {\bibinfo {title} {Nonlinear elastic behavior of graphene: \textit{Ab
  initio} calculations to continuum description},\ }\href
  {https://doi.org/10.1103/PhysRevB.80.205407} {\bibfield  {journal} {\bibinfo
  {journal} {Phys. Rev. B}\ }\textbf {\bibinfo {volume} {80}},\ \bibinfo
  {pages} {205407} (\bibinfo {year} {2009})}\BibitemShut {NoStop}%
\bibitem [{\citenamefont {Andrew}\ \emph {et~al.}(2012)\citenamefont {Andrew},
  \citenamefont {Mapasha}, \citenamefont {Ukpong},\ and\ \citenamefont
  {Chetty}}]{PhysRevB.85.125428}%
  \BibitemOpen
  \bibfield  {author} {\bibinfo {author} {\bibfnamefont {R.~C.}\ \bibnamefont
  {Andrew}}, \bibinfo {author} {\bibfnamefont {R.~E.}\ \bibnamefont {Mapasha}},
  \bibinfo {author} {\bibfnamefont {A.~M.}\ \bibnamefont {Ukpong}},\ and\
  \bibinfo {author} {\bibfnamefont {N.}~\bibnamefont {Chetty}},\ }\bibfield
  {title} {\bibinfo {title} {Mechanical properties of graphene and
  boronitrene},\ }\href {https://doi.org/10.1103/PhysRevB.85.125428} {\bibfield
   {journal} {\bibinfo  {journal} {Phys. Rev. B}\ }\textbf {\bibinfo {volume}
  {85}},\ \bibinfo {pages} {125428} (\bibinfo {year} {2012})}\BibitemShut
  {NoStop}%
\bibitem [{\citenamefont {Kang}\ \emph {et~al.}(2020)\citenamefont {Kang},
  \citenamefont {Ye}, \citenamefont {Fang}, \citenamefont {You}, \citenamefont
  {Levitan}, \citenamefont {Han}, \citenamefont {Facio}, \citenamefont
  {Jozwiak}, \citenamefont {Bostwick}, \citenamefont {Rotenberg}, \citenamefont
  {Chan}, \citenamefont {McDonald}, \citenamefont {Graf}, \citenamefont
  {Kaznatcheev}, \citenamefont {Vescovo}, \citenamefont {Bell}, \citenamefont
  {Kaxiras}, \citenamefont {van~den Brink}, \citenamefont {Richter},
  \citenamefont {Ghimire}, \citenamefont {Checkelsky},\ and\ \citenamefont
  {Comin}}]{Kang2019DiracFA}%
  \BibitemOpen
  \bibfield  {author} {\bibinfo {author} {\bibfnamefont {M.}~\bibnamefont
  {Kang}}, \bibinfo {author} {\bibfnamefont {L.}~\bibnamefont {Ye}}, \bibinfo
  {author} {\bibfnamefont {S.}~\bibnamefont {Fang}}, \bibinfo {author}
  {\bibfnamefont {J.-S.}\ \bibnamefont {You}}, \bibinfo {author} {\bibfnamefont
  {A.~L.}\ \bibnamefont {Levitan}}, \bibinfo {author} {\bibfnamefont
  {M.}~\bibnamefont {Han}}, \bibinfo {author} {\bibfnamefont {J.~I.}\
  \bibnamefont {Facio}}, \bibinfo {author} {\bibfnamefont {C.}~\bibnamefont
  {Jozwiak}}, \bibinfo {author} {\bibfnamefont {A.}~\bibnamefont {Bostwick}},
  \bibinfo {author} {\bibfnamefont {E.}~\bibnamefont {Rotenberg}}, \bibinfo
  {author} {\bibfnamefont {M.~K.}\ \bibnamefont {Chan}}, \bibinfo {author}
  {\bibfnamefont {R.~D.}\ \bibnamefont {McDonald}}, \bibinfo {author}
  {\bibfnamefont {D.~E.}\ \bibnamefont {Graf}}, \bibinfo {author}
  {\bibfnamefont {K.}~\bibnamefont {Kaznatcheev}}, \bibinfo {author}
  {\bibfnamefont {E.}~\bibnamefont {Vescovo}}, \bibinfo {author} {\bibfnamefont
  {D.~C.}\ \bibnamefont {Bell}}, \bibinfo {author} {\bibfnamefont
  {E.}~\bibnamefont {Kaxiras}}, \bibinfo {author} {\bibfnamefont
  {J.}~\bibnamefont {van~den Brink}}, \bibinfo {author} {\bibfnamefont
  {M.}~\bibnamefont {Richter}}, \bibinfo {author} {\bibfnamefont {M.~P.}\
  \bibnamefont {Ghimire}}, \bibinfo {author} {\bibfnamefont {J.~G.}\
  \bibnamefont {Checkelsky}},\ and\ \bibinfo {author} {\bibfnamefont
  {R.}~\bibnamefont {Comin}},\ }\bibfield  {title} {\bibinfo {title} {{D}irac
  fermions and flat bands in the ideal kagome metal {FeSn}},\ }\href
  {https://doi.org/10.1038/s41563-019-0531-0} {\bibfield  {journal} {\bibinfo
  {journal} {Nat. Mater.}\ }\textbf {\bibinfo {volume} {19}},\ \bibinfo {pages}
  {163} (\bibinfo {year} {2020})}\BibitemShut {NoStop}%
\bibitem [{\citenamefont {Ye}\ \emph {et~al.}(2018)\citenamefont {Ye},
  \citenamefont {Kang}, \citenamefont {Liu}, \citenamefont {von Cube},
  \citenamefont {Wicker}, \citenamefont {Suzuki}, \citenamefont {Jozwiak},
  \citenamefont {Bostwick}, \citenamefont {Rotenberg}, \citenamefont {Bell},
  \citenamefont {Fu}, \citenamefont {Comin},\ and\ \citenamefont
  {Checkelsky}}]{Ye2017MassiveDF}%
  \BibitemOpen
  \bibfield  {author} {\bibinfo {author} {\bibfnamefont {L.}~\bibnamefont
  {Ye}}, \bibinfo {author} {\bibfnamefont {M.}~\bibnamefont {Kang}}, \bibinfo
  {author} {\bibfnamefont {J.}~\bibnamefont {Liu}}, \bibinfo {author}
  {\bibfnamefont {F.}~\bibnamefont {von Cube}}, \bibinfo {author}
  {\bibfnamefont {C.~R.}\ \bibnamefont {Wicker}}, \bibinfo {author}
  {\bibfnamefont {T.}~\bibnamefont {Suzuki}}, \bibinfo {author} {\bibfnamefont
  {C.}~\bibnamefont {Jozwiak}}, \bibinfo {author} {\bibfnamefont
  {A.}~\bibnamefont {Bostwick}}, \bibinfo {author} {\bibfnamefont
  {E.}~\bibnamefont {Rotenberg}}, \bibinfo {author} {\bibfnamefont {D.~C.}\
  \bibnamefont {Bell}}, \bibinfo {author} {\bibfnamefont {L.}~\bibnamefont
  {Fu}}, \bibinfo {author} {\bibfnamefont {R.}~\bibnamefont {Comin}},\ and\
  \bibinfo {author} {\bibfnamefont {J.~G.}\ \bibnamefont {Checkelsky}},\
  }\bibfield  {title} {\bibinfo {title} {Massive {D}irac fermions in a
  ferromagnetic kagome metal},\ }\href {https://doi.org/10.1038/nature25987}
  {\bibfield  {journal} {\bibinfo  {journal} {Nature}\ }\textbf {\bibinfo
  {volume} {555}},\ \bibinfo {pages} {638} (\bibinfo {year}
  {2018})}\BibitemShut {NoStop}%
\bibitem [{\citenamefont {Liu}\ \emph {et~al.}(2018)\citenamefont {Liu},
  \citenamefont {Sun}, \citenamefont {Kumar}, \citenamefont {M{\"u}chler},
  \citenamefont {Sun}, \citenamefont {Jiao}, \citenamefont {Yang},
  \citenamefont {Liu}, \citenamefont {Liang}, \citenamefont {Xu}, \citenamefont
  {Kroder}, \citenamefont {S{\"u}ss}, \citenamefont {Borrmann}, \citenamefont
  {Shekhar}, \citenamefont {Wang}, \citenamefont {Xi}, \citenamefont {Wang},
  \citenamefont {Schnelle}, \citenamefont {Wirth}, \citenamefont {Chen},
  \citenamefont {Goennenwein},\ and\ \citenamefont {Felser}}]{Liu2017GiantAH}%
  \BibitemOpen
  \bibfield  {author} {\bibinfo {author} {\bibfnamefont {E.}~\bibnamefont
  {Liu}}, \bibinfo {author} {\bibfnamefont {Y.}~\bibnamefont {Sun}}, \bibinfo
  {author} {\bibfnamefont {N.}~\bibnamefont {Kumar}}, \bibinfo {author}
  {\bibfnamefont {L.}~\bibnamefont {M{\"u}chler}}, \bibinfo {author}
  {\bibfnamefont {A.}~\bibnamefont {Sun}}, \bibinfo {author} {\bibfnamefont
  {L.}~\bibnamefont {Jiao}}, \bibinfo {author} {\bibfnamefont {S.}~\bibnamefont
  {Yang}}, \bibinfo {author} {\bibfnamefont {D.}~\bibnamefont {Liu}}, \bibinfo
  {author} {\bibfnamefont {A.}~\bibnamefont {Liang}}, \bibinfo {author}
  {\bibfnamefont {Q.}~\bibnamefont {Xu}}, \bibinfo {author} {\bibfnamefont
  {J.}~\bibnamefont {Kroder}}, \bibinfo {author} {\bibfnamefont
  {V.}~\bibnamefont {S{\"u}ss}}, \bibinfo {author} {\bibfnamefont
  {H.}~\bibnamefont {Borrmann}}, \bibinfo {author} {\bibfnamefont
  {C.}~\bibnamefont {Shekhar}}, \bibinfo {author} {\bibfnamefont
  {Z.}~\bibnamefont {Wang}}, \bibinfo {author} {\bibfnamefont {C.}~\bibnamefont
  {Xi}}, \bibinfo {author} {\bibfnamefont {W.}~\bibnamefont {Wang}}, \bibinfo
  {author} {\bibfnamefont {W.}~\bibnamefont {Schnelle}}, \bibinfo {author}
  {\bibfnamefont {S.}~\bibnamefont {Wirth}}, \bibinfo {author} {\bibfnamefont
  {Y.}~\bibnamefont {Chen}}, \bibinfo {author} {\bibfnamefont {S.~T.}\
  \bibnamefont {Goennenwein}},\ and\ \bibinfo {author} {\bibfnamefont
  {C.}~\bibnamefont {Felser}},\ }\bibfield  {title} {\bibinfo {title} {Giant
  anomalous {H}all effect in a ferromagnetic kagom{\'e}-lattice semimetal},\
  }\href {https://doi.org/10.1038/s41567-018-0234-5} {\bibfield  {journal}
  {\bibinfo  {journal} {Nat. Phys.}\ }\textbf {\bibinfo {volume} {14}},\
  \bibinfo {pages} {1125} (\bibinfo {year} {2018})}\BibitemShut {NoStop}%
\bibitem [{\citenamefont {Wang}\ \emph
  {et~al.}(2013{\natexlab{b}})\citenamefont {Wang}, \citenamefont {Liu},\ and\
  \citenamefont {Liu}}]{WangOTI2013}%
  \BibitemOpen
  \bibfield  {author} {\bibinfo {author} {\bibfnamefont {Z.~F.}\ \bibnamefont
  {Wang}}, \bibinfo {author} {\bibfnamefont {Z.}~\bibnamefont {Liu}},\ and\
  \bibinfo {author} {\bibfnamefont {F.}~\bibnamefont {Liu}},\ }\bibfield
  {title} {\bibinfo {title} {Organic topological insulators in organometallic
  lattices},\ }\href {https://doi.org/10.1038/ncomms2451} {\bibfield  {journal}
  {\bibinfo  {journal} {Nat. Commun.}\ }\textbf {\bibinfo {volume} {4}},\
  \bibinfo {pages} {1471} (\bibinfo {year} {2013}{\natexlab{b}})}\BibitemShut
  {NoStop}%
\bibitem [{\citenamefont {Fang}\ \emph
  {et~al.}(2012{\natexlab{a}})\citenamefont {Fang}, \citenamefont {Gilbert},\
  and\ \citenamefont {Bernevig}}]{PhysRevB.86.115112}%
  \BibitemOpen
  \bibfield  {author} {\bibinfo {author} {\bibfnamefont {C.}~\bibnamefont
  {Fang}}, \bibinfo {author} {\bibfnamefont {M.~J.}\ \bibnamefont {Gilbert}},\
  and\ \bibinfo {author} {\bibfnamefont {B.~A.}\ \bibnamefont {Bernevig}},\
  }\bibfield  {title} {\bibinfo {title} {Bulk topological invariants in
  noninteracting point group symmetric insulators},\ }\href
  {https://doi.org/10.1103/PhysRevB.86.115112} {\bibfield  {journal} {\bibinfo
  {journal} {Phys. Rev. B}\ }\textbf {\bibinfo {volume} {86}},\ \bibinfo
  {pages} {115112} (\bibinfo {year} {2012}{\natexlab{a}})}\BibitemShut
  {NoStop}%
\bibitem [{\citenamefont {Fang}\ \emph
  {et~al.}(2012{\natexlab{b}})\citenamefont {Fang}, \citenamefont {Gilbert},
  \citenamefont {Dai},\ and\ \citenamefont
  {Bernevig}}]{PhysRevLett.108.266802}%
  \BibitemOpen
  \bibfield  {author} {\bibinfo {author} {\bibfnamefont {C.}~\bibnamefont
  {Fang}}, \bibinfo {author} {\bibfnamefont {M.~J.}\ \bibnamefont {Gilbert}},
  \bibinfo {author} {\bibfnamefont {X.}~\bibnamefont {Dai}},\ and\ \bibinfo
  {author} {\bibfnamefont {B.~A.}\ \bibnamefont {Bernevig}},\ }\bibfield
  {title} {\bibinfo {title} {Multi-{W}eyl topological semimetals stabilized by
  point group symmetry},\ }\href
  {https://doi.org/10.1103/PhysRevLett.108.266802} {\bibfield  {journal}
  {\bibinfo  {journal} {Phys. Rev. Lett.}\ }\textbf {\bibinfo {volume} {108}},\
  \bibinfo {pages} {266802} (\bibinfo {year} {2012}{\natexlab{b}})}\BibitemShut
  {NoStop}%
\bibitem [{\citenamefont {Zhang}\ \emph {et~al.}(2009)\citenamefont {Zhang},
  \citenamefont {Tang}, \citenamefont {Girit}, \citenamefont {Hao},
  \citenamefont {Martin}, \citenamefont {Zettl}, \citenamefont {Crommie},
  \citenamefont {Shen},\ and\ \citenamefont {Wang}}]{zhang2009direct}%
  \BibitemOpen
  \bibfield  {author} {\bibinfo {author} {\bibfnamefont {Y.}~\bibnamefont
  {Zhang}}, \bibinfo {author} {\bibfnamefont {T.-T.}\ \bibnamefont {Tang}},
  \bibinfo {author} {\bibfnamefont {C.}~\bibnamefont {Girit}}, \bibinfo
  {author} {\bibfnamefont {Z.}~\bibnamefont {Hao}}, \bibinfo {author}
  {\bibfnamefont {M.~C.}\ \bibnamefont {Martin}}, \bibinfo {author}
  {\bibfnamefont {A.}~\bibnamefont {Zettl}}, \bibinfo {author} {\bibfnamefont
  {M.~F.}\ \bibnamefont {Crommie}}, \bibinfo {author} {\bibfnamefont {Y.~R.}\
  \bibnamefont {Shen}},\ and\ \bibinfo {author} {\bibfnamefont
  {F.}~\bibnamefont {Wang}},\ }\bibfield  {title} {\bibinfo {title} {Direct
  observation of a widely tunable bandgap in bilayer graphene},\ }\href
  {https://doi.org/10.1038/nature08105} {\bibfield  {journal} {\bibinfo
  {journal} {Nature}\ }\textbf {\bibinfo {volume} {459}},\ \bibinfo {pages}
  {820} (\bibinfo {year} {2009})}\BibitemShut {NoStop}%
\bibitem [{\citenamefont {Taychatanapat}\ and\ \citenamefont
  {Jarillo-Herrero}(2010)}]{taychatanapat2010electronic}%
  \BibitemOpen
  \bibfield  {author} {\bibinfo {author} {\bibfnamefont {T.}~\bibnamefont
  {Taychatanapat}}\ and\ \bibinfo {author} {\bibfnamefont {P.}~\bibnamefont
  {Jarillo-Herrero}},\ }\bibfield  {title} {\bibinfo {title} {Electronic
  transport in dual-gated bilayer graphene at large displacement fields},\
  }\href {https://doi.org/10.1103/PhysRevLett.105.166601} {\bibfield  {journal}
  {\bibinfo  {journal} {Phys. Rev. Lett.}\ }\textbf {\bibinfo {volume} {105}},\
  \bibinfo {pages} {166601} (\bibinfo {year} {2010})}\BibitemShut {NoStop}%
\bibitem [{\citenamefont {Usui}\ \emph {et~al.}(2013)\citenamefont {Usui},
  \citenamefont {Donnelly}, \citenamefont {Logar}, \citenamefont {Sinclair},
  \citenamefont {Schoonman},\ and\ \citenamefont
  {Prinz}}]{usui2013approaching}%
  \BibitemOpen
  \bibfield  {author} {\bibinfo {author} {\bibfnamefont {T.}~\bibnamefont
  {Usui}}, \bibinfo {author} {\bibfnamefont {C.~A.}\ \bibnamefont {Donnelly}},
  \bibinfo {author} {\bibfnamefont {M.}~\bibnamefont {Logar}}, \bibinfo
  {author} {\bibfnamefont {R.}~\bibnamefont {Sinclair}}, \bibinfo {author}
  {\bibfnamefont {J.}~\bibnamefont {Schoonman}},\ and\ \bibinfo {author}
  {\bibfnamefont {F.~B.}\ \bibnamefont {Prinz}},\ }\bibfield  {title} {\bibinfo
  {title} {Approaching the limits of dielectric breakdown for {SiO$_2$} films
  deposited by plasma-enhanced atomic layer deposition},\ }\href
  {https://doi.org/10.1016/j.actamat.2013.09.003} {\bibfield  {journal}
  {\bibinfo  {journal} {Acta Mater.}\ }\textbf {\bibinfo {volume} {61}},\
  \bibinfo {pages} {7660} (\bibinfo {year} {2013})}\BibitemShut {NoStop}%
\bibitem [{\citenamefont {Hattori}\ \emph {et~al.}(2015)\citenamefont
  {Hattori}, \citenamefont {Taniguchi}, \citenamefont {Watanabe},\ and\
  \citenamefont {Nagashio}}]{hattori2015layer}%
  \BibitemOpen
  \bibfield  {author} {\bibinfo {author} {\bibfnamefont {Y.}~\bibnamefont
  {Hattori}}, \bibinfo {author} {\bibfnamefont {T.}~\bibnamefont {Taniguchi}},
  \bibinfo {author} {\bibfnamefont {K.}~\bibnamefont {Watanabe}},\ and\
  \bibinfo {author} {\bibfnamefont {K.}~\bibnamefont {Nagashio}},\ }\bibfield
  {title} {\bibinfo {title} {Layer-by-layer dielectric breakdown of hexagonal
  boron nitride},\ }\href {https://doi.org/10.1021/nn506645q} {\bibfield
  {journal} {\bibinfo  {journal} {ACS Nano}\ }\textbf {\bibinfo {volume} {9}},\
  \bibinfo {pages} {916} (\bibinfo {year} {2015})}\BibitemShut {NoStop}%
\end{thebibliography}%
\end{document}